\documentclass[conference]{IEEEtran}
\IEEEoverridecommandlockouts
\usepackage{amsmath,amssymb,amsfonts}
\usepackage{graphicx}
\usepackage{textcomp}
\usepackage{xcolor}
\usepackage{titlesec}
\usepackage{algpseudocode}
\def\BibTeX{{\rm B\kern-.05em{\sc i\kern-.025em b}\kern-.08em
    T\kern-.1667em\lower.7ex\hbox{E}\kern-.125emX}}

\usepackage[font=footnotesize,labelfont={bf}, figurename=Figure, tablename=Table]{caption}
\renewcommand{\thetable}{\arabic{table}}

\usepackage[activate={true,nocompatibility},final,tracking=true,kerning=true,spacing=true,factor=1100,stretch=10,shrink=10]{microtype}

\renewcommand{\thesection}{\arabic{section}}
\titleformat{\section}{\normalfont\Large\filcenter\bfseries}{}{0em}{}
\renewcommand{\thesubsection}{}
\titleformat{\subsection}{\normalfont\large\filcenter\bfseries}{\thesubsection}{0em}{}
\renewcommand{\thesubsubsection}{}
\titleformat{\subsubsection}
{\normalfont\normalsize\bfseries}{\thesubsubsection}{0.25em}{}
\titlespacing{\subsubsection}{0.25em}{.25em}{.25em}

\usepackage{xcolor}
\usepackage{placeins}
\usepackage{todonotes}

\usepackage{changes}
\usepackage[noabbrev,nameinlink, capitalise]{cleveref}
\crefname{appendix}{}{}

\usepackage[giveninits=true,style=nature,sorting=none,citestyle=numeric-comp,autocite = superscript, maxcitenames = 10]{biblatex}

\AtBeginBibliography{%
}

\DefineBibliographyStrings{english}{andothers = {\finalandcomma\ \upshape et al\adddot}}

\DeclareFieldFormat{labelnumberwidth}{#1\adddot}
\DeclareFieldFormat
  [article,book,incollection,inproceedings,patent,thesis,unpublished]
  {title}{#1\isdot}
\DeclareFieldFormat
  [article,book,incollection,inproceedings,patent,thesis,unpublished]
  {doi}{#1\isdot}
  
\DeclareFieldFormat
  [article,book,incollection,inproceedings,patent,thesis,unpublished]
  {year}{(#1)\isdot}

\DeclareFieldFormat
  [article,book,incollection,inproceedings,patent,thesis,unpublished]
  {volume}{#1, }
\DeclareFieldFormat
  [article,book,incollection,inproceedings,patent,thesis,unpublished]
  {journaltitle}{#1\ }
\DeclareFieldFormat
  [article,book,incollection,inproceedings,thesis,unpublished]
  {booktitle}{#1\ }

\DeclareFieldFormat
  [thesis]
  {volume}{#1.}

\DeclareFieldFormat
  [patent]
  {booktitle}{#1,\ }

\DeclareFieldFormat
  [misc]
  {title}{#1.\ }

\DeclareFieldFormat
  [article,book,incollection,inproceedings,patent,thesis,unpublished]
  {pages}{#1\isdot}

\DeclareListFormat
  [book]
  {publisher}{(#1)\isdot}
  
\DeclareBibliographyDriver{article}{%
  \printnames{author}%
  \newunit\newblock
  \printfield{year}%
  \newunit\newblock
  \printfield{title}%
  \newunit\newblock
  \printfield{journaltitle}%
  \printfield{volume}%
  \printfield{pages}
  \newunit
  \printfield{doi}
  \finentry}

\DeclareBibliographyDriver{inproceedings}{%
  \printnames{author}%
  \newunit\newblock
  \printfield{year}%
  \newunit\newblock
  \printfield{title}%
  \newunit\newblock
  \printfield{booktitle}%
  \printfield{volume}
  \printfield{pages}
  \newunit\newblock
  \printfield{doi}
  \finentry}   

\DeclareBibliographyDriver{patent}{%
  \printnames{author}%
  \newunit\newblock
  \printfield{year}%
  \newunit\newblock
  \printfield{title}%
  \newunit\newblock
  \printfield{booktitle}%
  \printfield{pages}
  \newunit\newblock
  \printfield{doi}
  \finentry}  

\DeclareBibliographyDriver{thesis}{%
  \printnames{author}%
  \newunit\newblock
  \printfield{year}%
  \newunit\newblock
  \printfield{title}%
  \newunit\newblock
  \printfield{booktitle}%
  \printfield{volume}
  \newunit\newblock
  \printfield{doi}
  \finentry}  

\DeclareBibliographyDriver{unpublished}{%
  \printnames{author}%
  \newunit\newblock
  \printfield{year}%
  \newunit\newblock
  \printfield{title}%
  \newunit\newblock
  \printfield{doi}
  \finentry}  
  
\DeclareBibliographyDriver{book}{%
  \printnames{author}%
  \newunit\newblock
  \printfield{year}%
  \newunit\newblock
  \printfield{title}%
  \
  \printlist{publisher}%
  \finentry}
\AtEveryBibitem{\clearfield{note}}
\AtEveryBibitem{\clearfield{language}} 
\renewcommand*{\bibfont}{\footnotesize}
\newcommand{\Es}{E_\mathrm{s}}
\newcommand{\xd}[1]{u^{x}_{#1}}
\newcommand{\yd}[1]{u^{y}_{#1}}
\newcommand{\phid}[1]{u^{\varphi}_{#1}}
\newcommand{\Fx}[1]{F^{x}_{#1}}
\newcommand{\Fy}[1]{F^{y}_{#1}}
\newcommand{\Mphi}[1]{M^{\varphi}_{#1}}

\newcommand{\citep}[1]{\cite{#1}}
\newcommand{\citet}[1]{\cite{#1}}
\renewcommand{\cite}[1]{\supercite{#1}}
\newcommand{\pr}{\nu^*}
\newcommand{\youngs}{E^*}
\newcommand{\dens}{\bar \rho}

\usepackage{nameref}
\begin{document}
% \linenumbers
\onecolumn

\title{Differentiable graph-structured models for inverse design of lattice materials}

\author{\IEEEauthorblockN{Dominik Dold$^{*,1}$, Derek Aranguren van Egmond$^{*,1}$}
\IEEEauthorblockA{\textit{$^1$ European Space Agency, Advanced Concepts Team,}
\IEEEauthorblockA{\textit{European Space Research and Technology Centre, 2201 AZ Noordwijk, South Holland, The Netherlands}}\\[-1mm]
\small{$^*$ Both authors contributed equally to this work.}}
}

\maketitle

\begin{abstract}
\noindent Architected materials possessing physico-chemical properties adaptable to disparate environmental conditions embody a disruptive new domain of materials science. Fueled by advances in digital design and fabrication, materials shaped into lattice topologies enable a degree of property customization not afforded to bulk materials. A promising venue for inspiration toward their design is in the irregular micro-architectures of nature. However, the immense design variability unlocked by such irregularity is challenging to probe analytically. Here, we propose a new computational approach using graph-based representation for regular and irregular lattice materials.
Our method uses differentiable message passing algorithms to calculate mechanical properties, therefore allowing automatic differentiation with surrogate derivatives to adjust both geometric structure and local attributes of individual lattice elements to achieve inversely designed materials with desired properties. 
We further introduce a graph neural network surrogate model for structural analysis at scale. 
The methodology is generalizable to any system representable as heterogeneous graphs.
\end{abstract}


\begin{center}
\textbf{Graphical abstract}\\[0.5mm]
\includegraphics[width=0.75\textwidth]{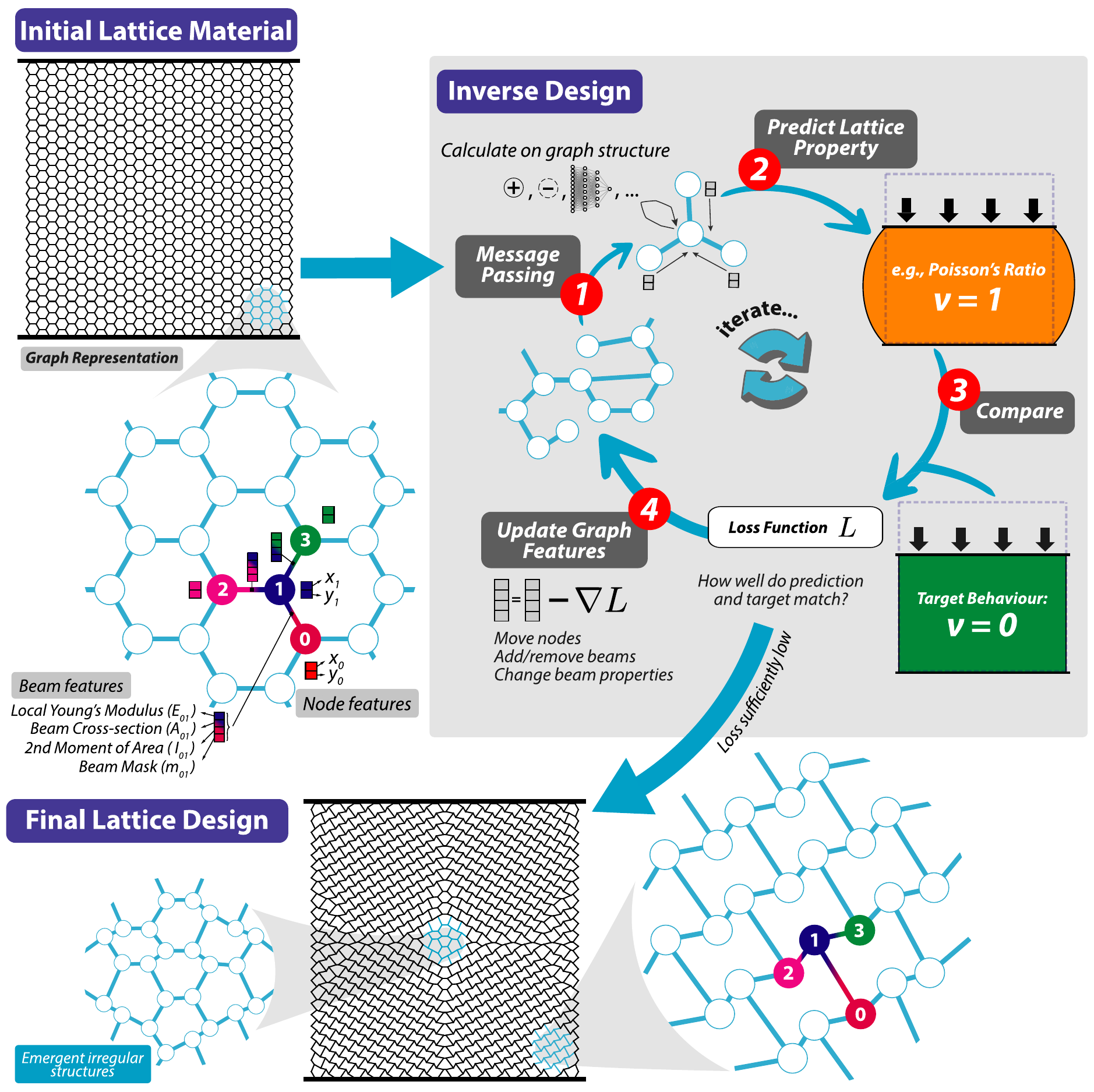}  
\end{center}

\FloatBarrier

\twocolumn

\section{Introduction}\label{sec:introduction}
Materials used to build future space infrastructure, especially those built directly on other planetary bodies, will be required to meet demanding conditions with environment-specific material properties, yet still be relatively easy to process and shape \cite{ghidini2018materials}. 
The constraints imposed by local planetary resources limit the palette of material composition that engineers can exploit to meet desired performance. Much like in nature, engineers will thus come to rely on the optimisation of a material’s topology in addition to its chemical makeup in order to achieve the desired properties under the limitations of the local milieu. Here on Earth, bone, plant stems, dragonfly wings, coral, and radiolarians \cite{gibson2010cellular} are just some examples of natural lattice materials that showcase how intricate architecture is used to achieve extreme mechanical performance with a limited choice of constituents \cite{wegst2015bioinspired}. Enhancements in e.g. strength, stiffness, impact toughness, fluid transport, and thermal insulation are all found while conserving light weight and minimizing mass transport. Moreover, an understanding of topological features often unlocks deformation modes and damage tolerant mechanisms not achieved by the bulk material alone.

Inspired by this, synthetic truss-based lattice materials - a subset of so-called Architected Materials -  comprise a highly active area of research in materials science. This activity is largely owed to the emergence of modern digital design and fabrication tools like 3D printing, or more formally, additive manufacturing. At the European Space Agency (ESA), capabilities for 3D printing lattice materials from a variety of space-relevant polymers, metals and in-situ planetary regoliths have been pioneered over the past decade \cite{makaya2022towards,mitchell2018additive}. The geometric freedom afforded by this technology, however, also creates an overwhelming design space of possible topological features, much like in nature. Human engineering has not been afforded the same evolutionary timescales as the natural optimisation pathways that drive biology's architectures. As such, the complete lattice material design space cannot be explored by engineering intuition alone.

An emergent stream in computational materials science demonstrates high potential for machine learning (ML)-driven design to aid engineers in the exploration of chemical and topological landscapes. Recent years have seen unprecedented demonstrations of physics-informed learning models capable of high fidelity material property predictions based on atomic-level interactions. Key among them are, e.g., the deployment of graph neural networks (GNNs) to demystify previously unsolved phase transition dynamics in glassy systems \cite{bapst2020unveiling}, and the use of generative adversarial networks (GANs) and variational autoencoders (VAEs) to predict complex molecular topologies in crystalline nanoporous materials like zeolites and metal-organic frameworks \cite{kim2020inverse, yao2021inverse}. 
 
Along with these developments, research into mechanical metamaterials has also pivoted towards ML as a means to augment the engineer's intuition with a data-driven geometric design language \cite{lee2023deep, challapalli2021inverse, guo2020semi, guo2021artificial, shen2022nature, zheng2021controllable, zheng2023deep, maurizi2022inverse, maurizi2022predicting, yang2022high}. Inverse design – the prescription of desired target properties and optimisation over various candidate microstructures –  has taken off as a vehicle for ML-aided simulations.  Whether supervised (data-driven) \cite{mozaffar2019deep, wang2020data} or unsupervised (physics-driven) \cite{bastek2022inverting}, these ML-based prediction models more accurately determine lattice mechanics, and dramatically accelerate the search for candidate architectures that meet target mechanical properties, forgoing computationally expensive  meshing required in finite element (FE) based simulations and conventional Topology Optimisation calculations \cite{gao2019topology}.

Conventionally these mechanical lattice materials are formed from periodically repeating unit cells, as this makes their properties addressable analytically and via numerical homogenization. Recent studies have pioneered deep neural networks toward inverse design of auxetic lattices, mapping subtle design variations in unit cell construction parameters to gains in total effective lattice properties via surrogate models of homogenization at much lower computational cost \cite{wang2020datastruct, liao2022deep}. While powerful, homogenization assumes uniformity of lattice properties throughout the global material, and does not account for, e.g., local imperfections, stress concentrations and edge-effects from partially constrained unit cells.

Contrary to periodic lattice materials used for lightweighting (such as hexagonal honeycombs), irregular architected materials – similar to those found in nature – have been shown to boast extraordinary damage tolerance \cite{aranguren2018designing}, anisotropic functional grading \cite{kumar2020inverse} and other surprising properties emerging from local defects and aperiodicity. Control over the most mechanically ‘beneficial’ features of these irregular tilings must happen at local defects and cannot be exerted analytically \cite{bonfanti2020automatic}, leaving a near-infinite space of heterogenous geometric combinations not easily modelled. 
For this reason, most modern synthetic lattice designs have been restricted to periodic structures, leaving out an ever-growing design space whose potential remains untapped. 

To tap into this design space, we propose a gradient-descent-based optimisation approach for inverse-designing emergent properties of irregular lattices. 
The core idea of our methodology is to represent lattice materials as heterogeneous graphs and perform computations directly on the topology of this graph using an operation called ‘message passing’.
This allows us to seamlessly link effective mechanical properties of the superstructure to local lattice elements (e.g. defects) in a differentiable way -- hence enabling the usage of gradient descent to iteratively apply targeted modifications to the lattice, reaching a design with desired properties after only tens to hundreds of iterations.

\begin{figure}[tb!]
    \centering    \includegraphics[width=\columnwidth]{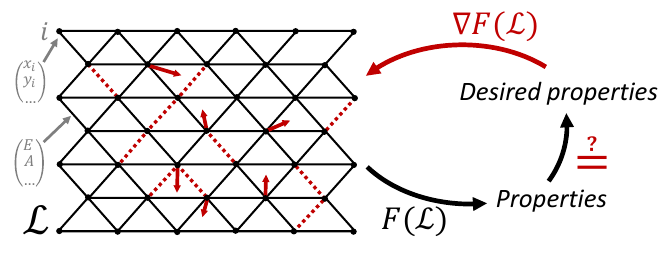}
	\caption{
	Schematic illustration of the proposed framework. A lattice material is represented as a graph $\mathcal{L}$ (left) which contains vector attributes both on the nodes (e.g. position) and edges (e.g. Young's modulus and beam cross-sectional area).
	A differential forward model $F$ takes the graph as input and predicts material properties.
	Comparing these predictions with desired properties, automatic differentiation is used to change the material (e.g., move nodes and remove or add beams) to better satisfy those properties.
	This process is repeated iteratively until a material with the desired properties is found.
	}\vspace{-5.8mm}	
	\label{fig:intro}
\end{figure}

More specifically, using message passing, we construct a differentiable forward model that predicts mechanical properties of lattice materials, which is then used in reverse through the application of automatic differentiation to change local lattice properties -- such as the cross-sectional area, parent material composition, and node positions of individual beams -- until a lattice structure with a set of desired global mechanical properties has been found (\cref{fig:intro}). 
Furthermore, by combining this inverse-design approach with a technique from computational neuroscience called ‘surrogate gradient’ \cite{neftci2019surrogate}, removal or addition of beams to the lattice using gradient information during inverse-design is enabled, opening up the possibility to quickly move through a huge space of candidate designs.

Conceptually, the introduced approach is inspired by recent work using GNNs to predict mechanical properties of periodic metamaterials. This was done by training said GNN only on the periodic unit cell structure \cite{ross2021using}. Our work extends this by enabling real-time structural feedback and inverse design for arbitrary aperiodic lattices, including complicated irregular topologies with local defects as found in natural materials.

In the following, we first provide a brief description of the methods used to model and characterize lattice materials.
Afterwards, we introduce the general idea of our proposed inverse design methodology, which we denote as `differentiable lattices' in the following.
Finally, two realizations of differentiable lattices will be presented, one based on an exact finite element method (requiring no training on data) and one using an approximate surrogate model, i.e., a GNN trained on simulated data. With both realizations, we demonstrate inverse design via gradient descent. Our initial focus is on the in-plane elastic properties of two-dimensional lattices for simplicity, though the methodology is easily transferred to three-dimensionally architectured designs.

\section{Results}\label{sec:results}
\subsection{2D lattice materials}\label{sec:methods}
Before introducing differentiable lattices, we briefly summarise how 2D lattice materials are modelled in the remainder of this work.
For more details on the used methods and concepts, see the experimental procedures. For details on simulations, see the supplemental experimental procedures.

\subsubsection{Notation and units}

In general, sets are denoted using calligraphic letters, vectors and matrices using bold letters and scalars using normal letters.
We summarise a lattice $\mathcal{L}$ as a tuple $\mathcal{L} = (\mathcal{X}, \mathcal{E}, \mathcal{A}_\text{E})$ consisting of node coordinates $\pmb r_i = (x_i, y_i) \in \mathcal{X}$ for each node $i$; the set of edges (beams) $\mathcal{E}$ between nodes, i.e., $(i,j) \in \mathcal{E}$ if an edge exists between nodes $i$ and $j$; and edge attributes $\pmb a_{ij} \in \mathcal{A}_\text{E}$.
For instance, lattice beams are characterized by their local Young's Modulus $E$, cross-sectional area $A$ and second moment of area $I$.
In principle, these can be chosen differently for each beam in the lattice. For simplicity, we choose one global value for all beams here.
Thus, for a homogeneous lattice, we have $\pmb a_{ij} = (E, A, I)\ \forall (i,j) \in \mathcal{E}$.

Distances between nodes are given by $\pmb r_{ij} = \pmb r_j - \pmb r_i$.
The length and orientation of a beam between nodes $i$ and $j$ is given by \cite{ochsner2018finite}
\begin{subequations}
\begin{align}
    L_{ij} &= \sqrt{(x_j - x_i)^2 + (y_j - y_i)^2}\,, \label{eq:L} \\
    s_{ij} &= \frac{y_j - y_i}{L_{ij}}\,, \label{eq:sin} \\
    c_{ij} &= \frac{x_j - x_i}{L_{ij}} \label{eq:cos} \,.
\end{align}
\end{subequations}
In the following, we abbreviate $L = L_{ij}$, $s = s_{ij}$ and $c = c_{ij}$. $s$ and $c$ here denote $\sin{\vartheta}$ and $\cos{\vartheta}$ respectively, where $\vartheta$ is the angle a beam makes with the lattice's base plane.
For simplicity, we assume that all lattices reside within a normalised bounding box with height $b_y$ and width $b_x$.
The proposed approach can be used for microscopic as well as macroscopic lattice structures. 
Here, we report results for a chosen set of material input paramaters, with $E = 2$ GPa, $b_x = b_y = 1$ cm and $A = 2\cdot 10^{-5}$cm$^2$.

\subsubsection{Modelling 2D lattice materials}\label{chap:FE}

To develop a general geometric model for lattice materials, we set out from the simplified case of a 2D lattice material, similar to a honeycomb sandwich panel used commonly in various engineering applications. To model in-plane mechanical properties we employ the direct stiffness method – a finite element matrix method derived from static analysis – to model elastic properties of our lattice material \cite{buhring2022elastic}. In it, a lattice is treated as a collection of connected beams, where each beam between nodes $i$ and $j$ is characterized by its stiffness matrix $\pmb{K}_{ij}$. 
The stiffness equation
\begin{equation}
    \pmb{K}_{ij} \begin{pmatrix}
    \pmb u_i \\
    \pmb u_j
    \end{pmatrix} = 
    \begin{pmatrix}
    \pmb F_i \\
    \pmb F_j
    \end{pmatrix} \,,
\end{equation}
allows us to calculate the reaction of the beam element to a given load.
Here, $\pmb u_i = (\xd{i}, \yd{i}, \phid{i})$ are the resulting node displacements due to external forces and moments $\pmb F_i = (\Fx{i}, \Fy{i}, \Mphi{i})$; and $\varphi$ characterises the resulting bending of beam elements.
In this work, we use generalized Euler-Bernoulli beam elements that can both deform along and perpendicular to their longitudal axis (see experimental procedures for details).

The global stiffness matrix $\pmb{G}$ for the whole lattice (i.e., more than two nodes) is constructed by summing up the contributions of each individual stiffness matrix for every node (see experimental procedures).
The final stiffness equation for a lattice material with $N$ nodes is then given by
\begin{equation}\label{eq:stiffnessEq}
     \pmb{G} \begin{pmatrix}
    \pmb u_0 \\
    \pmb u_1 \\
    ... \\
    \pmb u_{N-1}
    \end{pmatrix} = 
    \begin{pmatrix}
    \pmb F_0 \\
    \pmb F_1 \\
    ... \\
    \pmb F_{N-1}
    \end{pmatrix} \,.
\end{equation}\\

\subsubsection{Characterising 2D lattice materials}\label{chap:2Dprop}

A variety of mechanical in-plane properties are available to characterize the behaviour and functionality of 2D lattice materials. 
In this work, we focus on the relative density $\dens$, effective elastic modulus $\youngs$ and Poisson's ratio $\pr$. 
The asterisk denotes ``effective'' material properties for the entire lattice material, i.e., a global response, not just individual beams.
For non-isotropic materials, the elastic modulus and Poisson's ratio are direction-dependent.
Here, as a proof of concept, we restrict ourselves to these quantities measured along the vertical axis.
However, the presented results are applicable to properties measured along any axis, e.g., to inverse design the elastic modulus in horizontal and vertical direction at the same time.

\begin{figure*}[ht!]
    \centering
    \includegraphics[width=172mm]{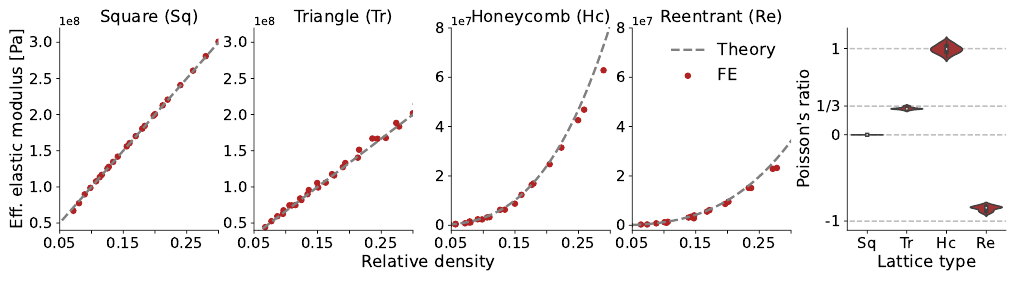}
	\caption{
	Comparison of analytical values (\cref{tab:rels}) and values obtained using our finite element method.
    We show the effective elastic modulus (left) and Poisson's ratio (right) for different tilings (square, equilateral triangle, honeycomb and reentrant honeycomb) and relative densities.
    To change the relative density, we repeated numerical experiments for lattices with different number of cells and beam cross-areas.
	}\vspace{-2mm}	
	\label{fig:analytical}
\end{figure*}

A detailed description of these mechanical properties and how they are obtained using the direct stiffness method is given in the experimental procedures.
For regular grids in 2D, $E^*$ and $\rho^*$ can be determined analytically \cite{Gibson_Ashby_1997, wang2005yield, lim2015auxetic} (see \cref{tab:rels}).
We use this to test our numerical approach, confirming that both analytical and numerical values agree for regular square, equilateral triangular, hexagonal honeycomb and reentrant honeycomb lattices with differing relative densities (\cref{fig:analytical}).
\subsection{Differentiable lattices}

In the following, we first introduce the general framework based on graph-based methods for inverse designing lattice materials.
Consequently, we demonstrate two different realizations of this framework: one using exact direct stiffness, and one using GNNs trained on experimental data.

\subsubsection{Representing lattices as graphs}

Lattices lend themselves to being modelled as graphs, with edges representing  beams and nodes representing the locations where beams connect with each other.
This allows an efficient and expressive description of lattices, where additional information such as node coordinates, beam cross-area, and beam Young's modulus can be encoded as node and edge features (i.e., real-valued vectors stored on nodes and edges) -- something that is not possible when representing lattices as, e.g., images.

\subsubsection{Calculating on graphs}

\begin{figure*}[ht!]
    \centering
    \includegraphics[width=172mm]{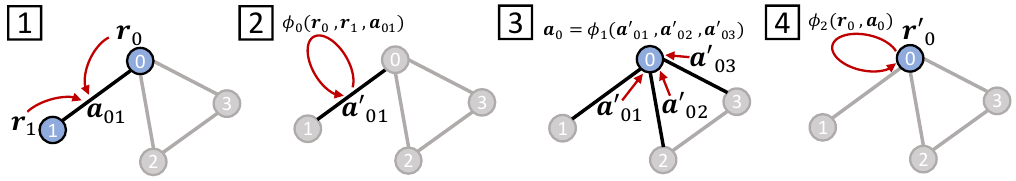}
	\caption{
	Illustration of the message passing steps. For clarity, we only show the operations on one edge, although in practice, these operations are performed on all edges in parallel. 
	(1) Node features $\pmb r_i$ are propagated along connecting edges.
	(2) On the edges, propagated node features as well as edge features are combined via a function $\phi_0$ to create new edge features. $\phi_0$ can take any shape (e.g., a neural network).
	(3) Edge features are reduced using a function $\phi_1$ that satisfies permutation invariance (i.e., the order of arguments does not matter) and is applicable to a varying number of arguments.
	(4) The node features are updated using the reduced edge features using a function $\phi_2$ (which, again, can take any shape).
	}\vspace{-4mm}	
	\label{fig:MP}
\end{figure*}
A recent and widely adopted approach of performing calculations on graph-structured data is message passing \cite{welling2016semi,gilmer2017neural,hamilton2017inductive}.
Message passing describes information flow between nodes that are directly connected with an edge, meaning that calculations are performed on edges with stored feature vectors that are locally available to that edge.
Usually, operations using message passing can be decomposed into two steps: a \textit{messaging} step and a \textit{reduction} step (\cref{fig:MP}).
In the messaging step, node features are sent along edges, where they are combined with edge features (and other node features) to calculate new edge features.
In the reduction step, newly calculated edge features are sent to neighbouring nodes and combined to form new node features.
These operations are performed on all edges and nodes in parallel, allowing efficient and scalable computations on the graph structure itself.
Most importantly, this realizes differentiable operations on the discrete structure of the graph, allowing us to utilize gradient-descent based optimization to, e.g., change initial features or even the structure of the graph itself to change the output of the calculation implemented by message passing.
This is enabled due to computations following the structure of the graph, i.e., all operations are differentiable, but the sequence of operations (i.e., how and which features are combined) is determined by the connectivity of the graph.

\subsubsection{Differentiable algorithms for inverse design}

We propose to use differentiable message passing algorithms for property prediction on lattices represented as graphs, which in turn can then be used to realize an iterative inverse design approach using gradient-descent based optimization.
In general, we denote by $F(\mathcal{L}, \mathcal{S})$ a function that takes node coordinates, graph edges and edge attributes $\mathcal{L} = (\mathcal{X}, \mathcal{E}, \mathcal{A}_\text{E})$
as input and returns one or several material properties of interest, e.g., effective elastic modulus, Poisson's ratio or simply the displacement of each node given a certain external distortion.
Optionally, $F$ can also take specific constraints $\mathcal{S}$ as input, describing which nodes in the lattice are forced or kept fixed.
To ease notation, we neglect writing $\mathcal{S}$ as an argument of $F$ in the following.
Although $F$ is still ambiguous here, it can take several shapes, as will be shown later.
Generally, $F$ is composed of several message passing steps, followed by pooling operations (combining node features) and usual differentiable operations such as neural networks.

For inverse design, we compare the predicted property $F(\mathcal{L})$ with a desired target value $\zeta$ (which, if several properties are predicted, takes the form of a vector).
How well prediction and target agree is measured using a loss function $L$, in our case a L1 loss
\begin{equation}
    L(F, \mathcal{L}, \zeta) = \| F(\mathcal{L}) - \zeta\| \,.
\end{equation}
This loss function is used to find a lattice material with the desired target properties by minimizing it using gradient descent, e.g., by iteratively changing the geometry of the lattice (adjusting node positions using $\nabla_{\pmb r_i}$ or removing/adding beams as described in the next subsection) or by changing the material properties of individual beams (i.e., changing edge features such as the cross-area of individual beams). 
In this work, we restrict ourselves to geometric changes only to find lattices with desired mechanical properties.
Gradient descent is implemented using automatic differentiation, which is readily available in current deep learning libraries such as Tensorflow and pyTorch.

\subsubsection{Masking edges}

To enable the removal or addition of beams in a lattice using gradient descent, we introduce an approach inspired by Ying et al.\cite{ying2019gnnexplainer} where each edge obtains an additional attribute: a mask value $m_{ij} \in \mathbb{R}$ that is used to decide whether a beam is realized in the lattice between nodes $i$ and $j$, i.e., $\pmb a_{ij} = (E, A, I, m_{ij})$.
In our case, the masking value is turned into a binary decision by applying the Heaviside step function $\theta(\cdot)$
\begin{equation}\label{eq:mask}
 m^\theta_{ij} = \theta\left(m_{ij}\right) = \begin{cases}
    1\,, & \text{if }  m_{ij} > 0 \,,\\
    0\,,              & \text{otherwise} \,,
\end{cases} 
\end{equation}
which is used to mask away the contribution of an edge during the reduction step -- as if it were not present in the lattice ($m^\theta_{ij} = 1$ -- beam exists; $ m^\theta_{ij} = 0$ -- beam does not exist).

In Ying et al.\cite{ying2019gnnexplainer}, masks are only used to remove edges from a graph, since enabling adding edges between all possible nodes would be computationally unfeasible.
However, in our case, a node can only be connected to a selected few other nodes in its local neighbourhood, since long beams spanning the whole material are not of interest to us.
Thus, when starting from, e.g., a triangular lattice, we can add additional beams to neighbouring nodes that are initially masked out, but can be added during the inverse design process.
To guarantee that we do not add crossing beams, we have to generate a list $C_{(i,j)}$ that contains, for each edge $(i,j)$, other edges that would physically cross it.
From this, a final mask value
\begin{equation}
    M_{ij} = m^\theta_{ij} \cdot \prod_{(n,m) \in C_{(i,j)}} \left(1 - m^\theta_{nm} \right) \,,
\end{equation}
is obtained, which basically unites the two conditions for a beam to be active in the lattice: its mask value has to be greater than $0$ and all other beams that would cross it have to be masked out.

During inverse design, both the list of crossing beams as well as the list of locally neighbouring nodes where beams could be introduced can be adjusted, enabling a complete geometric restructuring of the lattice material. 
In this work, for simplicity, we only update the list of crossing beams to ensure valid lattice designs with non-crossing beams.

\subsubsection{Surrogate gradients}

One problem remains: the decision function for masking, \cref{eq:mask}, has a Dirac delta distribution as its derivative, meaning that it vanishes everywhere except at the threshold, $m_{ij} = 0$.
This slows down optimization via gradient descent tremendously -- a problem that is well known in other areas such as computational neuroscience, where gradient-based learning for spiking neural networks faces the same problem.
However, recently, an approach called ``surrogate gradients'' \cite{neftci2019surrogate,zenke2021remarkable} has been introduced that enables robust gradient-based learning of spiking neural networks.

For learning to mask edges in a graph (or beams in a lattice), we apply the same trick: instead of using the Dirac delta function, we substitute it with a surrogate function with non-vanishing parts off the threshold.
A multitude of choices exist for surrogate functions.
Specifically for this work, we use a  mirrored Lorentz function $g$ as in Zenke et al.\cite{zenke2018superspike}:
\begin{equation}
    g\left(x\right) = \frac{1}{\left(\alpha \cdot |x| + 1 \right)^2} \,, 
\end{equation}
with $\alpha \in \mathbb{R}^+$ being a choosable hyperparameter and $|x|$ being the absolute value of $x$.
If not stated otherwise, we use $\alpha = 1$.

\subsection{Message passing finite element}\label{chap:mpfe}

As a first realization of $F$, we show how the direct stiffness method can be realized using message passing to form an end-to-end differentiable pipeline that returns exact mechanical properties given the graph representation of a lattice material.

\subsubsection{Model description}

The direct stiffness approach consists of several steps: 1) constructing the stiffness matrix for each beam, 2) adding masking to enable optimization of the beam connectivity, 3) combining those matrices into a global stiffness matrix describing the whole lattice, and 4) applying the experimental protocol for acquiring the desired mechanical property. In our framework, these steps take the following form:

\begin{enumerate}
    \item 
\textbf{\emph{Stiffness matrices}}: For an edge $(i,j)$, node features $\pmb r_i$ and $\pmb r_j$ (the ``messages'') are turned into edge features $L_{ij}$, $s_{ij}$ and $c_{ij}$, see \cref{eq:L,eq:sin,eq:cos}. Those edge features are sufficient to construct the stiffness matrix $\pmb{K}_{ij}$ on the edge, see \cref{eq:rod,eq:beam} in the experimental procedures.
\item 
\textbf{\emph{Masking}}: To mask a beam, the binary masking value $M_{ij}$ is multiplied to the stiffness matrix $\pmb{K}_{ij}$. 
Hence, if a beam is masked out, its contribution will not appear in the global stiffness matrix.
\item
\textbf{\emph{Global stiffness matrix}}: Constructing the global stiffness matrix is equivalent to a pooling operation that takes the features (masked stiffness matrix) of each edge and combines it into one global quantity valid for the whole graph (see \cref{fig:MPFE} for an illustration).
\item
\textbf{\emph{Properties}}: As discussed in the supplemental experimental procedures, to calculate the resulting deformation of the lattice given an external load, operations such as selecting parts of arrays and matrices, matrix-vector products, solving a system of linear equations and addition are required, which are all differentiable.
Similarly, the additional operations needed to obtain the effective elastic modulus and Poisson's ratio (e.g. linear regression) are differentiable as well.
For the relative density, we can simply sum up $L_{ij}$, as calculated using message passing in 1), over all edges $(i,j) \in \mathcal{E}$ with $M_{ij} > 0$ to obtain $L^*$ and subsequently $\dens$ (see \cref{eq:rhopoisson} in the experimental procedures).
\end{enumerate}
\begin{figure}[t!]
    \centering
    \includegraphics[width=\columnwidth]{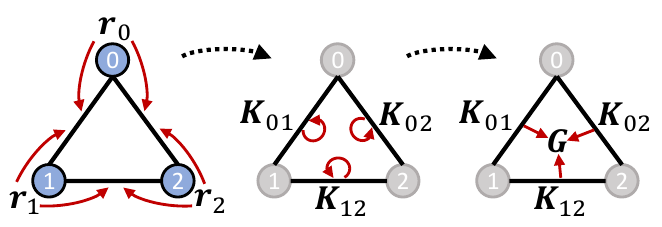}
	\caption{
	Illustration of message passing finite element. Using message passing on the graph representation of the lattice, first the stiffness matrix of each beam is computed (locally on each edge of the graph), from which the global stiffness matrix $G$ is then constructed (by pooling from all edges).
	}\vspace{-4mm}	
	\label{fig:MPFE}
\end{figure}

Thus, obtaining mechanical properties such as the relative density, resulting node displacements due to a load, effective elastic modulus and Poisson's ratio can be obtained in a fully differentiable framework, starting with message passing on the graph representation of the lattice, an edge-wise pooling operation and a series of ordinary differentiable operations, i.e., operations on data without graph structure, such as scalars, vectors and matrices.

The described forward model $F(\mathcal{L})$ specifically takes the following information from the graph $\mathcal{L}$ as input: the edge list $\mathcal{E}$, the masking values $\mathcal{M} = \{ m_{ij} \ | \ (i,j) \in \mathcal{E}\}$ and the node features $\mathcal{X} = \{\pmb r_i \ | \ 0 \leq i < N\}$ for a lattice with $N$ nodes, $F(\mathcal{L}) = F(\mathcal{X}, \mathcal{E}, \mathcal{M})$.
Since it is fully differentiable, it can be used to inverse design a lattice with desired properties using automatic differentiation -- both to change continuous properties (such as node coordinates) and discrete properties (such as beam existence).
In the following, we show this for two scenarios: obtaining a global target property such as a certain effective elastic modulus or Poisson's ratio, and obtaining a certain (functional) deformation given a loading scenario.

\subsubsection{Designing lattices with target properties}
\begin{figure*}[t!]
    \centering
    \includegraphics[width=\textwidth]{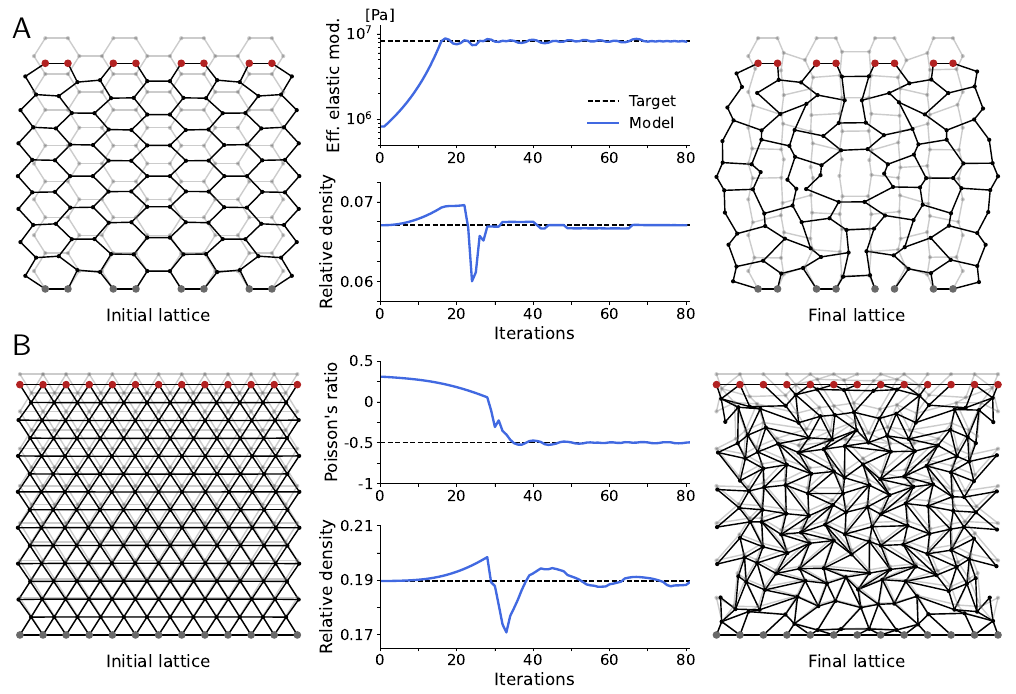}
	\caption{
	Inverse designing lattices to have desired mechanical properties. 
	Lattices without load are shown in light gray and with load in black.
	Top nodes (red) are forced, while bottom nodes (gray) are constrained to not move. In the middle plots, target values of properties are shown as dashed lines. 
    \textbf{(A)} Inverse designing a square lattice to have a higher effective elastic modulus while keeping the relative density constant. Deformations are magnified by a factor of $10$. 
    \textbf{(B)} Inverse designing a triangle lattice to have a negative Poisson's ratio while keeping the relative density constant. Deformations are not magnified.
	}\vspace{-4mm}	
	\label{fig:invDesign_Stiffness}
 \label{fig:invDesign_Poisson}
\end{figure*}

We demonstrate our approach with two examples:  starting with a regular honeycomb grid, we inverse design node positions and beam connectivity to acquire a lattice with an effective elastic modulus that is one order of magnitude higher than originally while keeping the relative density $\dens$ of the lattice constant (\cref{fig:invDesign_Stiffness}A).
In addition, we turn an initially regular triangular lattice with positive Poisson's ratio into a lattice with a negative Poisson's ratio of $\nu^\text{target} = -0.5$, again with the condition of keeping $\dens$ unchanged (\cref{fig:invDesign_Poisson}B).
In both cases, we start with all edges unmasked (i.e. beams that are originally not available cannot be added during inverse design) and the used loss function is a L1 loss with an additional regularization term for the relative density, see the supplemental experimental procedures for details.

In both cases, the required target values are achieved after a small number of iterations.
For the honeycomb lattice, the inverse design leads to a restructuring that is more akin to a square lattice, which matches our expectations as square grids have high axial stiffness.
To get a configuration with the same relative density as the initial lattice, beams that only weakly influence $\youngs$ are removed after the target elastic modulus has been reached.
For the triangular lattice, a configuration is found that, in general, promotes inward bending of elements during compression, resulting in a negative Poisson's ratio.
A peculiar feature are the arc-like structures on the bottom and top of the lattice that promote such inward movements.

\subsubsection{Designing lattices with target deformations} 
\begin{figure*}[t!]
    \centering
    \includegraphics[width=\textwidth]{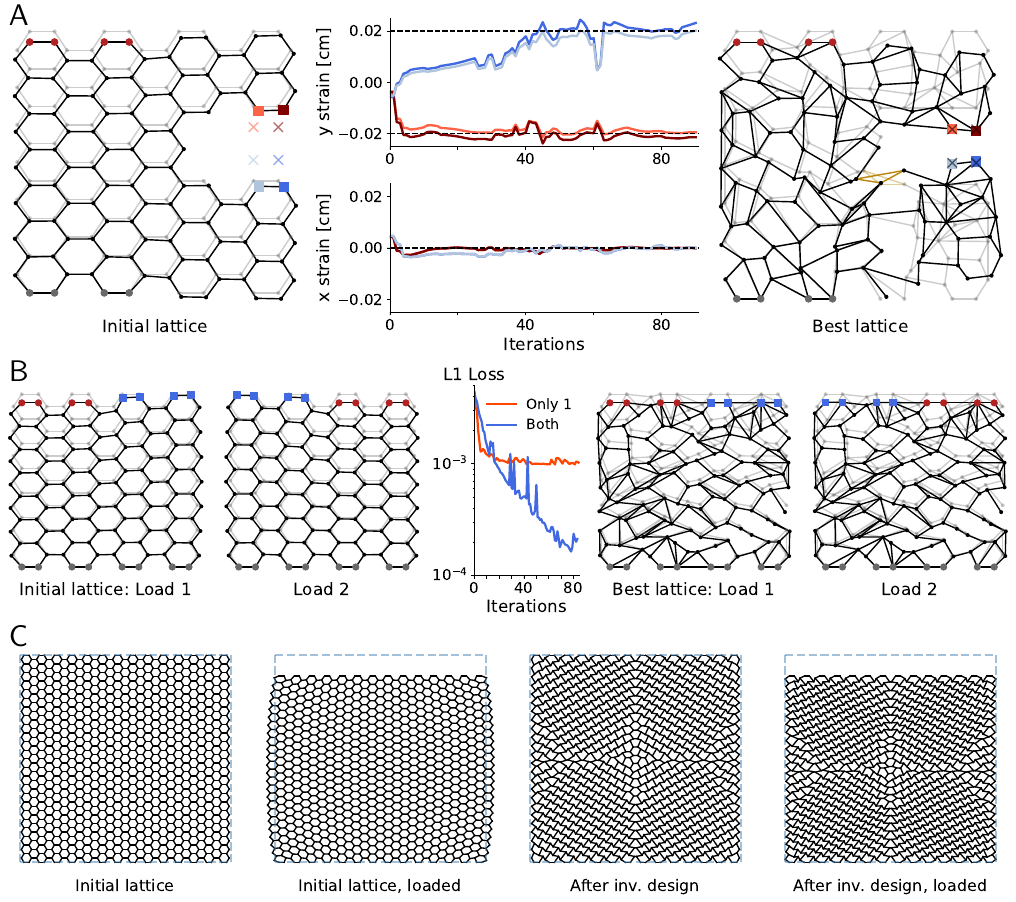}
	\caption{
	Inverse designing a lattice to yield certain deformation responses to loads. Colors and line styles as in \cref{fig:invDesign_Stiffness}. Nodes with target behaviour are shown as colored squares.
    \textbf{(A)} Inverse design of a grabbing tool. 
    Progress towards the target for all four nodes is shown in the middle plot, colored like the corresponding nodes (left and right, colored squares).
    In the lattice illustrations (left and right), target deformations are indicated by crosses. A lever-like structure that emerged during inverse design is highlighted in gold.
    Deformations are magnified by a factor of $4$. See also Video S1.
    \textbf{(B)} Inverse designing the response to two different load scenarios.
    As can be seen in the middle figure, if we only train on load scenario 1, the target behaviour for load scenario 2 is not reached.
    However, training on both scenarios will lead to a solution that satisfies both target behaviours.
    Deformations are magnified by a factor of $5$. See also \cref{fig:invDesign_FlatSI} and Video S2. \textbf{(C)} Inverse designing a large honeycomb lattice to have zero Poisson's ratio. Initially, loading the lattice leads to a bulging outwards. With our framework, a design is found that leads to no bulging, featuring local tiling motives resembling tilted reentrant honeycombs and triangles. See also \cref{fig:LargeAuxMagnified} and Video S3.
	}\vspace{-4mm}	
	\label{fig:invDesign_Grab}
    \label{fig:invDesign_Flat}
    \label{fig:Large}
\end{figure*}

In addition to global material properties, we can also inverse design the node displacements of the lattice to a given load.
In \cref{fig:invDesign_Grab}A, a lattice material that can perform a grabbing motion is found through inverse design.
We start with a hexagonal honeycomb lattice with a small cavity on the right side -- which will eventually become the grabbing part. To initiate the grabbing motion, the top left part of the lattice is compressed downwards while the bottom left part is kept in place. Initially, this leads to an outward bending of the right part of the lattice. Instead we turn this into an inward grabbing motion through inverse design.
To achieve this, we provide a target deformation in $y$ and $x$-direction (shown as crosses in \cref{fig:invDesign_Grab}A) for some of the nodes (colored squares) in the material's cavity.
As shown in \cref{fig:invDesign_Grab}A (middle), the four nodes converge towards their target behaviour after around 40 iterations.
The inverse design creates interesting functional structures, such as a lever-like arrangement (situated in the left bottom corner of the hole in the lattice, shown in gold in \cref{fig:invDesign_Grab}A) that pulls up the bottom-right part of the material when the lattice is compressed.

As a second example, we demonstrate in \cref{fig:invDesign_Flat}B that our approach can be used to design a lattice with several target behaviours. Specifically, we design a lattice structure that keeps a flat surface when only the top left (load 1) or top right half (load 2) is pushed downwards (with the other half free to move). 
As a target behaviour, we aim at keeping the top surface flat, i.e., the free moving part has to mimic the movement that would occur if the whole top surface was pushed downwards.
Although the target behaviour is only learned for one particular load strength, the acquired design is valid over a large range of external deformations, see \cref{fig:invDesign_FlatSI}.
For the scenarios shown in \cref{fig:invDesign_Grab,fig:invDesign_Flat}A,B, we allow new beams to be added between nodes that were initially not present in the lattice, drastically increasing the design space.

To show that our method can be scaled to much larger lattices, we design a honeycomb lattice composed of 635 cells to have no displacement in $x$-direction (for surface nodes) given a load in $y$-direction (\cref{fig:Large}C) -- which is equivalent to having a lattice material with $\nu^* = 0$.
Starting from a hexagonal honeycomb lattice, after only nine iterations a peculiar design is found using a motive resembling tilted reentrant honeycombs, producing the desired effect. We emphasize here that this motive is completely and autonomously emergent, with no prior conception of auxetic behaviour introduced to the model.
A magnified version of the design, featuring node positions, is shown in \cref{fig:LargeAuxMagnified}.

In general, the computational complexity of the exact model for $F$ is dominated by solving the stiffness equation (\cref{eq:stiffnessEq}), which scales with the cube of the number of cells (see \ref{si:tcompl} and \cref{fig:Tcompl} for details). Still, this is counter-balanced by the fact that gradient descent finds a suitable design typically within a few iterations -- making the approach viable even for larger lattices. To obtain results faster, a surrogate model for the finite element analysis can be used, i.e., trading accuracy for speed, such as GNNs trained on simulated or experimental data, where the computational complexity only scales linearly with the number of cells.

\subsection{Graph neural networks}\label{chap:gnn}

Instead of a finite element method, $F$ can also be an approximate model obtained using, e.g., machine learning. 
This is particularly useful when an analytical treatment is not feasible, computationally too slow, or only experimental data is available.
Recently, GNNs using convolutional operators \cite{welling2016semi,hamilton2017inductive,schlichtkrull2018modeling} have reached competitive performance on a variety of graph inference tasks such as link prediction and node classification, translating into applications such as molecule property prediction \cite{lu2019molecular}, mesh-based simulations \cite{pfaff2020learning}, modelling glassy systems \cite{bapst2020unveiling}, generalized neural algorithm learners \cite{ibarz2022generalist}, as well as material science and chemistry \cite{reiser2022graph}.
They are especially interesting due to their property of being able to deal with graphs that have a varying number of nodes as well as carry numerical features on nodes and edges.
Hence, in the following, we provide a proof of concept for using GNNs both for predicting the properties of lattice materials as well as inverse designing novel lattices.

\subsubsection{Dataset generation}

For training, evaluating and testing GNNs on predicting mechanical properties, we require an extensive dataset.
As a proof of concept, we generated simulated data using the direct stiffness method.
Lattices with different base tiling (square, equilateral triangle, hexagonal honeycomb and reentrant honeycomb) and deformations (random node displacements and beam/node removals) have been generated, with 4000, 200 and 1000 lattices of each tiling (train, validation and test data, respectively) -- i.e., in total 16000 training, 800 validation and 4000 test samples.
A detailed description of these perturbations can be found in the supplemental experimental procedures and \cref{SI:dataTable}. A visualisation of examples from the training dataset and dataset statistics are shown in \cref{fig:dataset} and \cref{fig:dstatistics}.

\subsubsection{Property prediction}

First, we train models that utilize message passing to predict in-plane material properties of 2D lattices.
We investigate two different GNN architectures: (i) GNNs based on the simple EdgeConv layer introduced in Wang et al.\cite{wang2019dynamic}, as well as a message passing neural network (MPNN) architecture used for molecule property prediction \cite{lu2019molecular}.
The model architectures are explained in detail in the experimental procedures.
For comparison, we further train a linear regression and CatBoost model \cite{prokhorenkova2018catboost} (i.e., a gradient-boosted tree) on handcrafted features extracted from the lattices such as cell density and relative density (see \cref{tab:tabular}).
As an alternative, focusing only on the geometry of the lattice, we also train a convolutional neural network (CNN) on image representations of the lattices.

To estimate the performance of the models, we report in \cref{tab:results} the root mean squared error (RMSE) calculated on the test set.
The best model during training is selected using the validation split.
In general, the CNN, CatBoost and MPNN perform similarly, clearly outperforming linear regression.
The best performance is reached by the EdgeConv GNN.
\begin{table}[!b]\renewcommand{\arraystretch}{1.2}
\begin{center}
\begin{tabular}{c || c | c }
% \cline{3-8}
\multicolumn{1}{c}{}& \multicolumn{1}{c}{Eff. elastic modulus} & \multicolumn{1}{c}{Poisson's ratio} \\
\hline\hline
Model & RMSE $[10^{-2}]$ & RMSE $[10^{-2}]$\\
\hline\hline
Lin. Regr. & $7.22$ & $23.00$ \\
CatBoost & $3.17$ & $10.89$ \\
CNN & $3.01$ & $10.53$ \\
EdgeConv & $2.61$ & $11.26$ \\
MPNN & $3.46$ & $10.65$ \\
\hline\hline
\end{tabular}
\vspace{1mm}
\caption{Experimental results for training prediction models.}
\label{tab:results}
\end{center}
\end{table}
More specifically, we found that the models learn to predict the effective elastic modulus from the lattice geometry (or tabular features) very well, while the Poisson's ratio is much harder to learn.
Especially for square lattices, we found that all models perform rather badly, although good performances are reached for reentrant honeycomb and hexagonal honeycomb lattices.
This is illustrated in \cref{fig:prediction}A, where we compare predicted and experimental values for the EdgeConv GNN.
These are promising results, especially since both GNN architectures are end-to-end differentiable and can thus be used as an approximate replacement of $F$ in the inverse design framework.

\subsubsection{Inverse design using graph neural networks}

\begin{figure*}[t!]
    \centering
    \includegraphics[width=172mm]{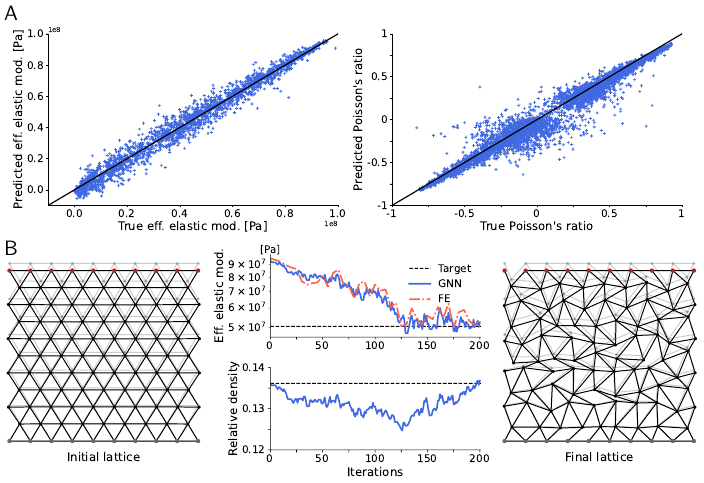}
	\caption{Experiments using graph neural networks as the forward model. \textbf{(A)} Model prediction vs. ground truth for the EdgeConv model. \textbf{(B)}
	Inverse design using a GNN as the forward model $F$. 
	A triangle lattice is adjusted to feature a reduced effective elastic modulus while its relative density has to be unchanged.
	Target values for the elastic modulus and relative density are shown as dashed lines.
	The predicted value by the GNN $F(\mathcal{L})$ is shown in blue (straight line), while the exact value obtained using finite element is shown in red (dashed dotted line).
	Colors and line styles as in \cref{fig:invDesign_Stiffness}.
	Deformations are magnified by a factor of $2$. See also \cref{fig:invDesignTrain}.
	}\vspace{-4mm}	
	\label{fig:invDesign_GNN}
 \label{fig:prediction}
\end{figure*}

To showcase the usage of GNNs for 2D lattice inverse design, we use an EdgeConv GNN trained to predict the effective elastic modulus as the forward model $F$.
To enable changing the beam structure of the lattice, we have to properly implement masking of edges in the message passing architecture of the GNN, which is explained in the experimental procedures.
In \cref{fig:invDesign_GNN}B, we show that inverse design is possible through a trained model, yielding a lattice design that has the desired target properties.
To ensure that the GNN did not simply return a lattice design from the training dataset, we show the two closest lattices from the training data in \cref{fig:invDesignTrain} (see \ref{SI:GNNdesign} for details).

\section{Discussion}\label{sec:discussion}
For inverse design, the exact forward model (message passing finite element) has the strong advantage that no training data is required at all.
In addition, the presented approach allows additional physical information of the lattice to be encoded as either node or edge features.
Thus, the scheme can be generalized to create completely heterogeneous lattices (each beam with different values of $E$ and $A$) to satisfy a given set of target properties.
In fact, due to the model being exact, a complete restructuring of the lattice geometry and its properties is possible without ever leaving the regime where the forward model is valid -- unless modifications are made that result in a free-hinged structure or disconnected material.
However, the model requires a full finite element implementation that -- depending on the size of the material and the degree of realism of the used finite element approach -- can be computationally inefficient and slow, having a cubic scaling relationship with respect to the number of cells in the lattice.
Still, as only a few iterations are required to find a single lattice design, we are confident that the approach scales to more complex applications and can be utilized as a design assistant to explore novel, irregular lattice materials.

An alternative to using an exact forward model is training a surrogate model, e.g., GNNs, which is computationally less demanding (linear scaling) than running a finite element simulation.
Furthermore, it can be applied to experimentally recorded data and a variety of (non-linear) mechanical properties.
In this work, we trained GNNs on predicting the effective elastic modulus $E^*$ and Poisson's ratio $\pr$ of lattice materials.
Although good performance is reached for predicting $E^*$, the model struggled with predicting $\pr$ accurately, especially for small absolute values of $\pr$.
To better understand the predictive power of GNNs in this application, a higher number of mechanical properties should be investigated in future work, as well as alternative GNN architectures that are more tailored to the problem of lattice material property prediction.
Moreover, we only used lattices with a predefined base tiling and small range of cells.
In future work, a larger dataset that covers a vast regime of lattice topologies (e.g. based on Voronoi grids) could strongly benefit the training process.

Major downsides of using GNNs is that a lot of training data is required -- more than might be accessible from experimental observations.
In this case, a transfer learning approach might be used, where the model is first trained with simulated data and then fine-tuned, or partially retrained to predict novel properties, using experimental data.
Finally, another possibility for reducing the amount of training data required as well as increase the performance and robustness of the model is to use a hybrid approach that integrates part of the analytical model in the GNN, basically molding knowledge about the underlying physics into the GNN architecture.
To allow beams to be added or removed during inverse design, edge mask values have to be added to the GNN, details of which strongly depend on the choice of GNN architecture.
For instance, in this work we showed how masking can be integrated in a GNN architecture using EdgeConv layers.

Different from inverse design using an exact, analytical forward model (e.g. \cref{fig:invDesign_Stiffness,fig:invDesign_Flat}), the optimization process is noisier when using a surrogate model.
Furthermore, we found that it works less reliably the further one moves away from the value regime covered during training of the model.
In fact, the design loop can get stuck in configurations of the lattice that are not realistic (e.g., disconnected components) without being able to recover from it on its own.
Although this can be accounted for in the inverse design framework, for instance by not allowing changes to the geometry that result in unphysical lattice materials, this is a clear downside of the approach compared to using an exact model.
Nevertheless, especially for larger lattices, inverse design with GNNs is extremely fast (\cref{fig:Tcompl}). Thus, if sufficient data is available to train a model, it can be used to quickly generate candidate designs, which then only need to be refined for a few iterations with the exact -- but much slower -- message passing finite element model, drastically speeding up the inverse design.

To mask in and out beams in the lattice design during inverse design, we used a technique called surrogate gradient. Without surrogate gradients, we found that only continuous properties of the lattice, such as node positions, are changed during gradient descent -- which is not too surprising, since the gradient of the threshold function is a Dirac delta distribution, which vanishes everywhere except at $0$, and thus mask values of beams are only rarely changed. In contrast, with surrogate gradients, beams were continuously added or removed during inverse design, enabling a discrete restructuring of the lattice.

When using this method, there is an ambiguity on what kind of function to use as the surrogate gradient. 
However, it has recently been observed that learning with surrogate gradients is robust to different shapes of the surrogate function -- at least for training spiking neural networks \cite{zenke2021remarkable}. In this work, we only used one type of surrogate function, but with adjustable width $\alpha$ -- which we found to not drastically change performance.

We used a masking scheme here that incorporates information about crossing beams in the inverse design framework. Although we decided to simply mask out all crossing beams, in principle, a scheme could be developed that samples -- from a group of crossing beams -- a ``winning'' beam. For instance, the Gumbel-Softmax trick \cite{JanGuPoo17}, which enables sampling from categorical distributions compatible with automatic differentiation, could be used to select the winning beam based on their edge mask values.

Although we only focused on 2D lattices here, the introduced method is extendable to more complex structures and materials. For instance, 3D truss-based lattices can be encoded in the same way as 2D lattices: node coordinates (in 3D) and an adjacency matrix describe the structure itself, while stiffness matrices are again calculated for each beam element using message passing.
In fact, our method is naturally compatible with any truss-based lattice and we are additionally confident that it can be generalized to more complex lattice types in the future, such as sheet- and shell-based materials. We envision this achieved e.g., by representing the mesh of local elements (instead of beams) as a heterogeneous graph, and using message passing between unit elements to model their interactions. An analogous approach has shown promise for property prediction from the grain microstructures of polycrystalline metallic alloys \cite{dai2021graph}.

% \subsection{Conclusion}\label{sec:conclusion}
To conclude, we present a framework that utilizes differentiable graph representations of lattices to perform both property prediction and inverse design.
The main focus lies hereby on using message passing algorithms, which perform calculations directly on the geometric structure of the lattice's graph representation and allow the modification of local material properties and beam connectivity using automatic differentiation to optimize global properties of the lattice.
We show that finite element methods can be realized using message passing, or GNNs trained on simulated data can act as surrogate models thereof.
This yields an efficient and expressive way of both describing and parametrizing lattices as well as modelling their behaviour mathematically.

Our approach constitutes an important step towards enabling automatic inverse design of irregularly structured lattice materials.
Moreover, it opens up a new set of tools developed in the graph machine learning literature, such as the GNNExplainer method, for analyzing both regular and irregular lattices.
We hope that this will spark new ideas for representing 3D-printable materials and lead to a wealth of novel approaches and tools that assist practitioners in designing new (multi)-functional materials.
An intriguing example of a technology that is highly synergistic with our approach is the recently introduced mechanical neural network\cite{lee2022mechanical} (MNN) -- a physical, experimental lattice structure constructed with programmable stiffness beam elements. Similar to how analogue implementations of neural networks have been trained \cite{schmuker2014neuromorphic,esser2016,schmitt2017neuromorphic,kungl2019accelerated}, MNNs could be trained ‘in-the-loop’ using our method, i.e., with forward passes being done on the physical device, while backward calculations (i.e. error backpropagation) for adjusting the stiffness of individual beams are done using our model on an edge device.
This way, MNNs could be trained in real-time to create adaptive and smart structures in the real world.

Finally, we would like to stress that the approach introduced in this work is not limited to describing lattices, but is applicable to any system that constitutes of a graph representation and a forward model that predicts its properties -- thus allowing automatic differentiation to change the structure of the input graph until it satisfies a set of desired properties.

\section*{Experimental procedures}

\subsection*{Generalized Euler-Bernoulli beam elements}

The stiffness matrix $\pmb{K}_{ij}$ of the generalized beam element is obtained by combining the stiffness matrices of rod elements $\pmb{K}^\mathrm{rod}_{ij}$ and Euler-Bernoulli beam elements $\pmb{K}^\mathrm{EB}_{ij}$,
\begin{equation}
    \pmb{K}_{ij} = \pmb{K}^\mathrm{rod}_{ij} + \pmb{K}^\mathrm{EB}_{ij} \,.
\end{equation}
Rod elements are used here to model the deformation of lattice elements along their longitudal axis.
Their corresponding stiffness matrix is given by \cite{ochsner2018finite}
\begin{equation}\label{eq:rod}
    \pmb{K}^\mathrm{rod}_{ij} = \frac{E A}{L}
\begin{pmatrix}
c^2 & c s & 0 & -c^2 & -c s & 0 \\
c s & s^2 & 0 &  -c s & -s^2 & 0 \\
0 & 0 & 0 & 0 & 0 & 0 \\
-c^2 & -c s & 0 & c^2 & c s & 0 \\
-c s & -s^2 & 0 & c s & s^2 & 0 \\
0 & 0 & 0 & 0 & 0 & 0
\end{pmatrix} \,.
\end{equation}
Euler-Bernoulli beam elements model the bending of beam elements where the beam length is much larger than the characteristic dimension of the cross section, for which the stiffness matrix is given by \cite{ochsner2018finite}
\begin{align}
    &\pmb{K}^\mathrm{EB}_{ij} = \frac{E I}{L^3} \cdot \label{eq:beam} \\
    & \begin{pmatrix}
    12 s^2 & -12 s c & -6 L s & -12 s^2 & 12 s c & -6 L s \\
    -12 s c & 12 c^2 & 6 L c & 12 s c & - 12 c^2 & 6 L c \\
    -6 L s & 6 L c  & 4 L^2 & 6 L s & -6 L c & 2 L^2 \\
    -12 s^2 &  12 s c & 6 L s & 12 s^2 & -12 s c & 6 L s \\
    12 s c & -12 c^2 & -6 L c & -12 s c & 12 c^2 & -6 L c \\
    -6 L s &  6 L c  & 2 L^2 & 6 L s &  -6 L c & 4 L^2
    \end{pmatrix} \nonumber \,.
\end{align}
We assume beam elements with a square-shaped cross-sectional area, which corresponds to $I=\frac{bh^3}{12} = \frac{t^4}{12} = \frac{A^2}{12}$, where $b, h$ denote cross-section depth and height, respectfully. In a square beam, these are both equal to beam thickness\cite{megson2019structural}, $t$.

\subsection*{Global stiffness matrix}

The global stiffness matrix $\pmb{G}$ can be constructed iteratively: first, start with $\pmb{G}$ being the zero-matrix (i.e., all elements are zero). 
Then, for each beam connecting two nodes $i$ and $j$ in the lattice, $\pmb{G}$ is updated as follows (with $\pmb{k} = \pmb{K}_{ij}$ here):
\begin{subequations}
\begin{align}
    \widetilde{\pmb{G}}_{ii} &\gets \widetilde{\pmb{G}}_{ii} + \widetilde{\pmb{k}}_{00} \,, \\
    \widetilde{\pmb{G}}_{ij} &\gets \widetilde{\pmb{G}}_{ij} + \widetilde{\pmb{k}}_{01}\,, \\
    \widetilde{\pmb{G}}_{ji} &\gets \widetilde{\pmb{G}}_{ji} + \widetilde{\pmb{k}}_{10}\,, \\
    \widetilde{\pmb{G}}_{jj} &\gets \widetilde{\pmb{G}}_{jj} + \widetilde{\pmb{k}}_{11}\,.
\end{align}
\end{subequations}
The index notation $a$:$b$ denotes the range of integers from $a$ to $b$, i.e., $\pmb{K}_{0:3}$ is the sub-matrix of $\pmb{K}$ consisting only of its first three rows.
We further introduced the specific index notation $\widetilde{\pmb{K}}_{ij} = \pmb{K}_{3\cdot i:3\cdot i+3,3\cdot j:3\cdot j+3}$, where the sub-selection is applied to both rows and columns.

\subsection{Mechanical in-plane properties}

\subsubsection*{Relative density}
The most influential design parameter on the in-plane mechanical properties of a cellular lattice material is its relative density\cite{Gibson_Ashby_1997}, $\dens = \frac{\rho^*}{\rho_\text{s}}$. 
This is defined as the ratio between the lattice's density  and the density of the parent solid material, $\rho_\text{s}$. 
For most practical scenarios, this ratio is equal to the material volume fraction contained within a known bounding box, following the relation: 
\begin{equation}
    \dens = \frac{\rho^*}{\rho_\mathrm{s}} = \frac{V_\mathrm{s}}{V_\mathrm{tot}} = \frac{L^* \cdot t}{b_x \cdot b_y}\,, \label{eq:rhopoisson}
\end{equation}
where $L^* = \sum_{(i,j) \in \mathcal{E}} L_{ij}$ is the sum of all beam lengths, $t = \sqrt{A}$ the beam thickness (and width) and $b_y$ and $b_x$ the height and width of the bounding box, respectively.

In cases where the bounding box is not known or considered, the relative density of common periodic lattices can also be determined analytically. 
\Cref{tab:rels} lists the relative density as a function of beam thickness, $t$, and regular beam length, $L$ for various periodic regular tilings. 
Note that contributions from material at nodal beam intersections are assumed negligible. The analytical relationships of \Cref{tab:rels} are only applicable for sufficiently slender beams when\cite{Gibson_Ashby_1997} $\dens < 0.2$. Otherwise, stress distribution within nodes and the emergence of axial shear effects in thicker beams cause the model to break down at higher relative densities. More robustly, Meza et al.\cite{meza2017reexamining} confirmed the classical predictions apply if strut dimensions fall within the regime $(t/l)\lesssim 0.05$.

To ensure our generalized beam element model yields viable elastic properties, we limit all our experiments to a regime of relative densities between $0.05 < \dens \le 0.19$, contingent on the varying number of cells and beam thicknesses in each observation. In most cases, our beams also satisfy the slenderness criterion \cite{meza2017reexamining}. 

\subsubsection*{Effective elastic modulus}\label{sec:E}

The effective elastic modulus $\youngs$ describes how strongly a material resists to externally-induced deformations.
In general, $\youngs$ is given by the slope of the linear regime of a material's stress-strain curve.

\begin{figure}[t!]
    \centering
    \includegraphics[width=\columnwidth]{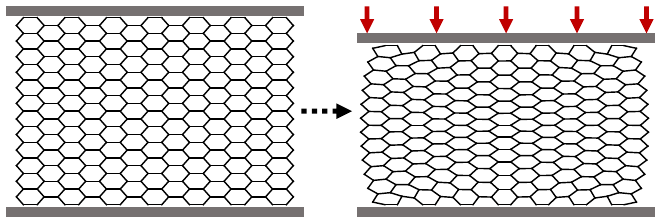}
	\caption{
	Experimental setup for determining both the effective elastic modulus and the Poisson's ratio of a lattice material along the loading direction. The material -- here a lattice with honeycomb tiling -- is placed between two plates. By pressing the top plate downwards while keeping the bottom plate unmoved, the material gets compressed. With this approach, stress-strain as well as strain-strain curves are derived, from which the above-mentioned material properties are calculated.
	}\vspace{-4mm}	
	\label{fig:exp}
\end{figure}

We use the following experimental setup to determine $\youngs$ for arbitrary -- including regular as well as irregular -- 2D lattice materials.
First, the material is glued between two plates in $y$-direction, i.e., we have a top and a bottom plate (\cref{fig:exp}).
To obtain the stress-strain curve, the top plate is then iteratively pushed downwards to force the top nodes of the lattice to move. Bottom nodes are constrained in place.
This yields different strains, i.e., displacements in y-direction.
The stress is then obtained by measuring what force the material is applying on the top plate in response to the induced displacements, and dividing by the cross-sectional area of the lattice's bounding box. 
After collecting several stress values for increasing strains, the effective elastic modulus is given by the slope of the resulting stress-strain curve.
How this experimental setup translates into simulations in detail is described in the supplemental experimental procedures.
In general, it requires constructing the global stiffness matrix $G$, applying constraints and deformations, solving the stiffness equation for $\pmb u_i$ and updating the node coordinates $\pmb r_i$ for all nodes $i$ -- which has to be repeated several times to record the stress-strain curve.

\subsubsection*{Poisson's ratio}\label{sec:rho}

The Poisson's ratio measures how the width of a material changes due to a forced compression of its height. 
For instance, many materials will widen when compressed, which corresponds to a positive Poisson's ratio.
However, so-called auxetic materials do the opposite: when compressed, their width is reduced as well, which corresponds to a negative Poisson's ratio.
If the width of the material does not change at all, its Poisson's ratio is $0$.

The Poisson's ratio can be obtained with a similar experimental setup as the effective elastic modulus, just that we measure the change in width due to a strain in y-direction.
Hence, the Poisson's ratio is given by the slope of a strain-strain curve, as explained in detail in the supplemental experimental procedures.
The width change is calculated by taking the difference of the mean $\xd{r}$ value of all nodes on the outer right surface and the mean $\xd{l}$ value of all nodes on the outer left surface of the material.

\subsection*{Graph neural networks}\label{SI:GNN}

We investigate graph neural networks that work directly on the graph structure of our lattices, using only the information contained in $\mathcal{L}$.

Here, we use the $\mathrm{EdgeConv}$ model \citep{wang2019dynamic}, where the node features $\pmb r_i$ are updated as follows:
\begin{equation}
    \pmb r_i^{(l+1)} = \mathrm{max}_{j \in \mathcal{N}_i} \left(\pmb{W}\,  \big(\pmb r^{(l)}_j - \pmb r^{(l)}_i \big) + \pmb{W}_0\, \pmb r^{(l)}_i \right) \,, \label{si:edgeconv}
\end{equation}
with $\pmb r_i^{(0)} = \pmb r_i$, $\mathcal{N}_i$ is the set containing all nodes connecting to node $i$, and $\pmb{W}$ and $\pmb{W}_0$ are matrices.
The full model consists of several $\mathrm{EdgeConv}$, followed by a dense deep neural network $\Phi$ that returns a property prediction for each node. 
The final prediction is then obtained by averaging over all nodes.

In addition, we investigate the $\mathrm{MPNN}$ model that has been proposed for molecule property prediction \citep{lu2019molecular}.
For this model, we use the distance vector between nodes $\pmb r_{ij}$, the length of the beam $L_{ij}$ as well as the orientation $c_{ij}$ as edge features $\pmb e_{ij} = \left(\pmb r_{ij}, L_{ij}, c_{ij}\right)$.
Node features are first preprocessed using a multi-layer neural network $\phi_\mathrm{r}$.
They are update using a $\mathrm{NNConv}$ layer and a gated recurrent unit ($\mathrm{GRU}$)
\begin{align}
    &\tilde{\pmb r}_i = \pmb r_i^{(l)} + \mathrm{mean}\{ \phi_\mathrm{e}(\pmb e_{ij}) \cdot \pmb r_j^{(l)},\ j \in \mathcal{N}_i\}  \,, \\
    &\left(\pmb r_i^{(l+1)},\, \pmb h_i^{(l+1)}\right) = \mathrm{GRU}\left(\tilde{\pmb r}_i,\, \pmb h_i^{(l)}\right) \,,
\end{align}
with $\pmb r_i^{(0)} = \pmb{h}_i^{(0)} =  \phi_\mathrm{r}\left(\pmb r_i\right)$ and where $\phi_\mathrm{e}$ is a neural network that takes the edge features as input and returns a matrix.
This step is repeated $N$ times, after which a graph embedding $\pmb r_\mathrm{g}$ is obtained by pooling over all nodes
\begin{equation}
    \pmb r_\mathrm{g} = \mathrm{pool}\left(\left\{\pmb r_0^{(N)},\ \ldots \ , \pmb r_{|\mathcal{C}|}^{(N)} \right\} \right) \,.
\end{equation}
For pooling, we use the Set2Set operator.
From $\pmb r_g$, a prediction is obtained through a final multi-layer neural network $\phi_\mathrm{p}$.

\subsection*{Masking EdgeConv}\label{SI:maskGNN}

For EdgeConv, message passing yields the following edge features for each edge $(j,i)$
\begin{equation}
    E_{ji} = \left(\pmb{W}\,  \big(\pmb r^{(l)}_j - \pmb r^{(l)}_i \big) + \pmb{W}_0\, \pmb r^{(l)}_i \right) \,.
\end{equation}
Masking is done as follows:
\begin{equation}
    E_{ji} \gets E_{ji} \cdot M_{ji} + \text{min}_{n,m} \left(E_{nm} \right) \cdot (1 - M_{nm}) \,.
\end{equation}
Node features are updated by choosing the maximum value (element-wise) over all neighbouring edges
\begin{equation}
    \pmb r_i^{(l+1)} = \mathrm{max}_{j \in \mathcal{N}_i} \left(E_{ji}\right) \,.
\end{equation}
Hence, if an edge is masked, its value is not picked by the $\mathrm{max}$ operation and it appears as if the edge does not exist in the graph.
The masking also guarantees that the chosen maximum value cannot be larger than the minimum value of the edge features, always guaranteeing that the masked value is not chosen by accident.
If the graph is bidirectional, both forward and backward edge between two nodes are masked with the same mask value, i.e., $m_{ji} \equiv m_{ij}$, resulting in $M_{ji} \equiv M_{ij}$.
Self-connections $(i,i)$ in the graph are not masked. This also guarantees that masking still works if nodes become disconnected from the remaining graph.

\subsection*{Data and code availability}
\noindent This publication is accompanied with an extensive Python module (based on pyTorch) for analysing and designing 2D lattices called \textit{pyLattice2D} (see \ref{SI:pyLattice2D}), which is publicly available on gitlab \cite{pylattice}. An archived version can be found under the following DOI: 10.5281/zenodo.8239350.
It contains the experiments performed in this work, as well as many convenience functions, e.g., for the comfortable generation of various 2D lattice geometries.
Default parameters as well as details for simulations can be found in the supplemental experimental procedures.

\section*{Acknowledgments}\label{sec:ack}
% \addcontentsline{toc}{section}{Acknowledgment}
We would like to thank Jai Grover and Elissa Ross for helpful and stimulating discussions.
We further thank the reviewers for their helpful feedback, and our colleagues at ESA’s Advanced Concepts Team for their ongoing support.
Both authors acknowledge support through the European Space Agency fellowship and young graduate trainee programs. 

\section*{Author contributions}
% \addcontentsline{toc}{section}{Author contributions}
Both authors designed the theoretical and experimental aspects of this study.
DD created the pyLattice2D package based on initial code by DAvE.
Both authors contributed to the pyLattice2D code.
DD performed the simulations.
Both authors wrote the manuscript based on an initial draft by DD.

\section*{Declaration of interests}
% \addcontentsline{toc}{section}{Declaration of interests}
Work was performed while both authors were employed with the European Space Agency. At the time of publication, DAvE's affiliation is the National Research Council Canada, Advanced Materials Research Facility, Mississauga, Canada.\\

\printbibliography
\addcontentsline{toc}{section}{References}

\onecolumn

\renewcommand{\appendixname}{Supplemental Information}
\captionsetup{list=no}
\begin{center}
\LARGE{\textbf{Supplemental Information}}\\[10pt]
\end{center}
\renewcommand{\thesection}{\arabic{section}}  
\renewcommand{\thefigure}{S\arabic{figure}}
\renewcommand{\theequation}{S\arabic{equation}}
\setcounter{equation}{0}
\setcounter{figure}{0}
\setcounter{subsection}{0}
\renewcommand{\thetable}{S\arabic{table}}
\setcounter{table}{0}
\addtocontents{toc}{\protect\setcounter{tocdepth}{0}}
\renewcommand{\thesubsection}{Note S\arabic{subsection}}
\titleformat{\subsection}{\normalfont\normalsize\bfseries}{\thesubsection :}{1em}{}

\begin{refsection}
\fontsize{10}{12}\selectfont
\noindent \textbf{\Large{Supplemental experimental procedures}}\\

% \clearpage
\noindent \textbf{\large{Determining mechanical properties using direct stiffness}}\\

\noindent \textbf{Calculating the effective elastic modulus}\\
For simplicity, we denote by $\mathcal{T}$ the set containing all indices of top nodes (i.e., all nodes forming the top surface of the material) and $\mathcal{B}$ the set containing the indices of bottom nodes.

\begin{enumerate}
    \item First, construct the global stiffness matrix $\pmb{G}$.
    \item A single external displacement $(0, \delta)$ of the top nodes is introduced by setting the constraints $\xd{i} = 0$ and $\yd{i} = \delta \ \forall i \in \mathcal{T}$.
    \item This displacement leads to forces acting on all remaining nodes
    \begin{equation}
        \pmb f_\text{ext} = - \pmb{G}\ \pmb u_\text{ext} \,,
    \end{equation}
    where $\pmb u_\text{ext} \in \mathbb{R}^{3N}$ is zero everywhere except for the external displacements, i.e., $u_{\text{ext}, 3\cdot i+1} = \yd{i}$ if $i \in \mathcal{T}$. 
    \item The ``glueing'' is imitated by enforcing the constraints $\xd{i} = 0$ and $\yd{i} = 0 \ \forall i \in \mathcal{B}$. Both the constraints for bottom and top nodes are realized by removing the corresponding rows and columns in the stiffness matrix and the corresponding rows in the external force vector (i.e., all rows (and columns) $3\cdot i$ and $3\cdot i+1$ $\forall i \in \mathcal{T} \cup \mathcal{B}$.), resulting in their reduced versions $\bar{\pmb{G}}$ and $\bar{\pmb f}_\text{ext}$.
    \item From the reduced stiffness matrix and force vector, we get the displacement of all unconstrained nodes $\bar{\pmb  u}$ by solving the following system of linear equations
    \begin{equation}
        \bar{\pmb{G}} \bar{\pmb u} = \bar{\pmb f}_\text{ext} \,,
    \end{equation}
    for instance, by using a linear (and differentiable) solver such as LU decomposition.
    \item From $\bar{\pmb u}$ and the constraints for top and bottom nodes, the displacement vector $\pmb u$ for all nodes can be constructed.
    \item From this, we can calculate the full force vector $\pmb f = \pmb{G} \pmb u$.
    \item The total force of the top nodes pushing upwards is then given by
    \begin{equation}
        f_\mathrm{react} = -\sum_{i \in \mathcal{T}} f_{3\cdot i+1}\,,
    \end{equation}
    which is turned into a stress by normalizing with the cross-sectional area of the lattice in y-direction, $b_x \cdot \sqrt{A}$,
    \begin{equation}
        \mathrm{stress} = \frac{f_\mathrm{react}}{b_x \cdot \sqrt{A}}\,.
    \end{equation}
    \item Finally, we update the node positions of the lattice to their new equilibrium positions $(x_i + \xd{i}, y_i + \yd{i})$.
\end{enumerate}
To collect stress values for increasing strain, the whole process is repeated several times. For instance, after two iterations, the total stress is given by the sum of the stresses obtained in both iterations, and the total strain is given by $2 \cdot \delta$. 
The effective elastic modulus is then obtained via linear regression on the collected total stress values $\pmb \sigma$ and total strain values $\pmb \epsilon$,
\begin{equation}
    \youngs = \frac{\sum_k \sigma_k}{\sum_k \epsilon_k} \,,
\end{equation}
where $k$ sums over all iterations.
This approach can be easily generalized to arbitrary scenarios by defining the sets $\mathcal{T}$ and $\mathcal{B}$ as well as the constraints and external displacements differently. 

Related to ``Characterising 2D lattice materials'' in the main text.\\

\noindent \textbf{Calculating the Poisson's ratio}\\
To obtain the Poisson's ratio, the same steps as for the effective elastic modulus are performed. However, instead of the stress, the mean width change $\bar{\epsilon}_k$ is calculated in each iteration $k$
\begin{align}
    \Delta R &= \frac{1}{|\mathcal{RS}|} \sum_{i \in \mathcal{RS}} \xd{i} \,, \\
    \Delta L &= \frac{1}{|\mathcal{LS}|} \sum_{i \in \mathcal{LS}} \xd{i} \,, \\
    \bar{\epsilon}_k &= \Delta R - \Delta L \,,
\end{align}
where $\mathcal{RS}$ is a set containing the indices of all outer-right nodes (i.e., forming the right surface of the material) and $\mathcal{LS}$ the indices of outer-left nodes. We only use outer nodes that are unconstrained for this calculation.
$|\mathcal{RS}|$ denotes the number of elements in $\mathcal{RS}$.
For deformations, we neglect the iterations index here to increase the readability.
The Poisson's ratio is then obtained using linear regression
\begin{equation}
    \pr = \frac{\sum_k \bar{\epsilon}_k}{\sum_k \epsilon_k} \,.
\end{equation}

\noindent Related to ``Characterising 2D lattice materials'' in the main text.\\

\noindent \textbf{\large{Simulation details}}\\

\noindent \textbf{Default simulation parameters}\\
During inverse design, we optimize both the node coordinates $\pmb r_i$ and mask values $m_{ij}$.
However, we only change the coordinates of nodes that are inside of the material, i.e., the outer surface of the material is kept unchanged.
For coordinates, we use the learning rate $\gamma_{\pmb r} = 0.001$ and for the edge mask $\gamma_{m} = 0.01$.
If not stated otherwise, we use $\alpha = 1$, $E = 2$ GPa, $b_x = b_y = 1$ cm and $A = 2\cdot 10^{-5}$cm$^2$ in all simulations.
The optimized parameters are not regularized.

For training GNNs, we use a batch size of $200$, a learning rate of $10^{-3}$ (Adam optimizer), weight regularization strength $10^{-6}$ and a mean squared error loss function.
For all trained models, we normalized the values of the effective elastic modulus by first subtracting the minimum value of the training set, and then dividing by the maximum value of the (minimum-shifted) training set. 

All simulations ran on an AMD Ryzen 9 5900HS and a NVIDIA GeForce RTX 3060 (Laptop).\\

\noindent \textbf{Inverse design of effective elastic modulus}\\
In this case, the exact forward model $F$ provides the effective elastic modulus of a given lattice.
To determine $\youngs$, $F$ performs $10$ iterations of direct stiffness with $\delta = 0.001$, leading to a total compression in height of $0.01$ (i.e. 1\%).
We choose the loss function
\begin{equation}
   L_E = \frac{\| F(\mathcal{X}, \mathcal{E}, \mathcal{M}) - E^\text{target}\|}{E_0} + \beta \| \dens - \dens_0\| \,,
\end{equation}
where $E_0$ is the initial effective elastic modulus of the lattice, $E^\text{target} = 10\cdot E_0$ the target value, $\dens$ the relative density as obtained using message passing and $\dens_0$ the initial relative density.
$\beta$ is a hyperparameter that we choose to be $\beta = 10$.

Masking values are initialized as $0.2$ for all edges.
No additional edges (beyond regular honeycomb connectivity) are added to the lattice setup.
The learning rate for mask values is reduced by a factor of $10$ after $40$ iterations to guarantee convergence. 

Related to \cref{fig:invDesign_Stiffness}A.\\

\noindent \textbf{Inverse design of Poisson's ratio}\\
In this case, the exact forward model $F$ provides the Poisson's ratio of a given lattice.
To determine $\pr$, $F$ performs $40$ iterations of direct stiffness with $\delta = \frac{0.01}{40}$, leading to a total compression in height of $0.01$ (i.e. 1\%).
We choose the loss function
\begin{equation}
    L_\nu = \| F(\mathcal{X}, \mathcal{E}, \mathcal{M}) - \nu^\text{target}\| + \beta \| \dens - \dens_0\| \,,
\end{equation}
where $\nu^\text{target} = -0.5$ is the target Poisson's ratio.
All other parameters are as in ``Inverse design of effective elastic modulus'' in the supplemental experimental procedures.
In addition, the list of crossing beams is recalculated every $10$ iterations to ensure a valid lattice material after inverse design has finished. 

Related to \cref{fig:invDesign_Stiffness}B.\\

\noindent \textbf{Inverse design of grabber}\\
Different from the previous experiments, here the deformation of each node is determined using $F$, which performs $10$ iterations of direct stiffness with $\delta = \frac{0.01}{10}$.
For the right outer nodes in the small cavity (\cref{fig:invDesign_Grab}, square-shaped nodes in blue and red), we set as a target that they do not move in $x$-direction, but move either upwards (lower row) or downwards (upper row) in $y$-direction by $0.02$.
Performance is evaluated using an L1 loss again, and we set $\alpha = 100$.

In this experiment, we allow new edges to be added to the lattice that have originally not been part of the honeycomb tiling.
Before starting the inverse design loop, we therefore add new edges (that are masked out initially) to the graph: for each node, edges to neighbouring nodes within a radial distance of $0.2$ are added to the graph with a probability of $0.3$.
Original honeycomb beams $(i,j)$ are initialized with mask value $m_{ij} = 0.25$, while all other (newly added) beams $(l,p)$ are initialized with mask value $m_{lp} = 0$.
During inverse design, we also apply the following to promote removing unnecessary beams from the lattice, as well as solutions that are further away from the initial honeycomb structure:
\begin{itemize}
    \item Every five iterations, the $10$\% of the active edges (i.e., $\{(i,j) \ | \ M_{ij} > 0\}$) with the lowest mask value are masked out by setting their mask value to $-0.2$. 
    \item Every five iterations, the expressions for $M_{ij}$ are updated by newly checking which beams in the lattice cross. This is done to avoid solutions with crossing beams.
\end{itemize}

Related to \cref{fig:invDesign_Grab}A.\\

\noindent \textbf{Inverse design of flat surface}\\
In this experiment, there are two loading scenarios $\mathcal{S}_1$ and $\mathcal{S}_2$, which we provide as an additional input to the forward model $F$.
More specifically, in the first scenario the four left nodes on the top of the surface are moved downwards, while in the second scenario the four right nodes on the top are moved.
In both scenarios, all bottom nodes are kept fixed and the target is to have a flat surface, i.e., the remaining four top nodes have as a target to not move in $x$-direction and move the same amount in $y$-direction as the forced nodes.

For training, we again use the L1 loss -- however, now it is the sum of the individual losses for both scenarios.
Training is done as in ``Inverse design of grabber'' in the supplemental experimental procedures, with the only difference being that random masking every five iterations is stopped after 20 iterations. 

Related to \cref{fig:invDesign_Grab}B.\\

\noindent \textbf{Inverse design of a large lattice}\\
The load is only applied in one iteration, with a deformation in $y$-direction of $\delta = \frac{0.02}{1}$.
For the outer left and outer right nodes, the target during inverse design is to have no displacement in $x$-direction.
As before, we use an L1 loss and initialize the masking values as $0.2$ for all edges.
The lattice deformations shown in the main paper (before and after inverse design) are applied in $50$ increments, i.e., $\delta = \frac{0.1}{50}$.

Related to \cref{fig:invDesign_Grab}C.\\

\noindent \textbf{Dataset generation}\\
For the dataset, we chose the number of layers $N_\text{l}$ in a material using a certain base tiling such that in the end, all lattices were made of approximately the same amount of cells $N_\text{c}$ (since all lattices have to fit as best as possible into the unit box, an exact match is impossible). The deformations are governed by hyperparameters $\Delta$, $D_\text{n}$ and $D_\text{e}$, and we denote by $U(x)$ the uniform distribution over the interval $[0,x]$ ($x \in \mathbb{R}^+)$ and $I(N)$ the uniform distribution on integers in the interval $[0,N]$ ($N \in \mathbb{N}$).

To deform a single lattice, maximum values for the deformations are obtained by sampling from random distributions:
\begin{subequations}
\begin{align}
A_\delta &\sim U(1) \,, \\
\partial &\sim U(\Delta) \,, \\
d_\text{e} &\sim U(D_\text{e}) \,, \\
d_\text{n} &\sim I(D_\text{n}) \,.
\end{align}
\end{subequations}
The deformations are then applied as follows:
\begin{enumerate}
    \item Select a fraction $A_\delta$ of nodes randomly and shift them by a random amount $U(\partial)-0.5$ both in x and y direction (for both directions, the amount is determined independently).
    \item Select a fraction $d_\text{e}$ of edges randomly and remove them from the edge list (i.e., remove beams from the lattice).
    \item Select $d_\text{n}$ nodes randomly and remove them as well as all edges connecting to them.
\end{enumerate}
The used values are listed in \cref{SI:dataTable}. 
For training, validation and test set, different random seeds were used.
This way, a variety of different lattices is generated, with some featuring heavy deformations while others only have small (or localized) deformations.
To calculate the effective elastic modulus and Poisson's ratio, we performed $30$ iterations of direct stiffness with $\delta = \frac{0.02}{30}$ on each lattice.
A brief summary of the distribution of these values is shown in \cref{fig:dstatistics}.

Related to \cref{fig:invDesign_GNN} and \cref{tab:results}.\\

\noindent \textbf{Tabular models}\\
For comparison, we consider both linear regression and gradient-boosted trees (CatBoost).
Such models work best with engineered features that summarise the characteristics of a lattice $\mathcal{L}$.
In addition to the graph representation of the lattice, we also use its image representation $\pmb{\mathcal{I}}$ here, which is given by a matrix $\pmb{\mathcal{I}} \in \{0,1\}^{c_0} \times \{0,1\}^{c_1}$, with $c_0 \times c_1 = 339 \times 459$ being the image resolution and a value of $1$ indicating the existence of lattice material.
Here, we use the features listed in \cref{SI:dgeneration}.

Related to \cref{fig:invDesign_GNN}A and \cref{tab:results}.\\

\noindent \textbf{Training machine learning models}\\
We train separate models for the effective elastic modulus and the Poisson's ratio (although in principle, similar performances are reached when training at least the GNN models to predict both properties).
For the GNN models, we choose a graph representation with bidirectional edges.
In addition, we add a self-connection to each node.

The EdgeConv model consists of three EdgeConv layers (see \cref{si:edgeconv}) with $200$ hidden neurons each. The deep neural network consists of three layers with $[400, 200, 1]$ neurons.

The specification of the MPNN model can be found in pyLattice2D (in \verb|models/MPNN/networks| the class \verb|LatticeNNConv|) with parameters \verb|hid_nfeat| set to 15 and \verb|num_message_passing| set to 3.

For the CNN, we use a batchsize of 50 and weight regularization of $10^{-5}$. Training data is randomly flipped horizontally and vertically.
The specifications of the CNN architecture can be found in pyLattice2D (in \verb|models/MPNN/networks| the class \verb|CNN|).

Related to \cref{fig:invDesign_GNN} and \cref{tab:results}.\\

\noindent \textbf{Inverse design using GNNs}\\
For inverse design, we trained an EdgeConv model with $150$ hidden neurons per EdgeConv layer, otherwise the architecture is the same as in ``Training machine learning models'' in the supplemental experimental procedures.
For the inverse design loop, we use again an L1 loss:
\begin{equation}
    L_\nu = \| F(\mathcal{X}, \mathcal{E}, \mathcal{M}) - E^\text{target}\| + \beta \| \dens - \dens_0\| \,,
\end{equation}
with $\beta = 100$ and $F$ now the trained EdgeConv model.
Only original beams of the triangle grid can be added or removed during training.
We initialize the mask randomly with values $m_{ij} \sim U(0.2)$, which was necessary to avoid that the model removes too many beams at once during the first few iterations.
For coordinates, we use the learning rate $\gamma_{\pmb r} = 0.0001$ and for the edge mask $\gamma_{m} = 0.001$.
All parameters are regularized (weight decay) with strength $10^{-6}$.

For comparison, we use an exact forward model (FE in \cref{fig:invDesign_GNN}B) which uses $10$ iterations of direct stiffness with $\delta = 0.002$ to determine $E^*$.

Related to \cref{fig:invDesign_GNN}B.\\

\noindent \textbf{\Large{Supplemental notes}}\\

\subsection{Computational complexity}\label{si:tcompl}

\begin{figure}[ht!]
    \centering
    \includegraphics[width=0.45\columnwidth]{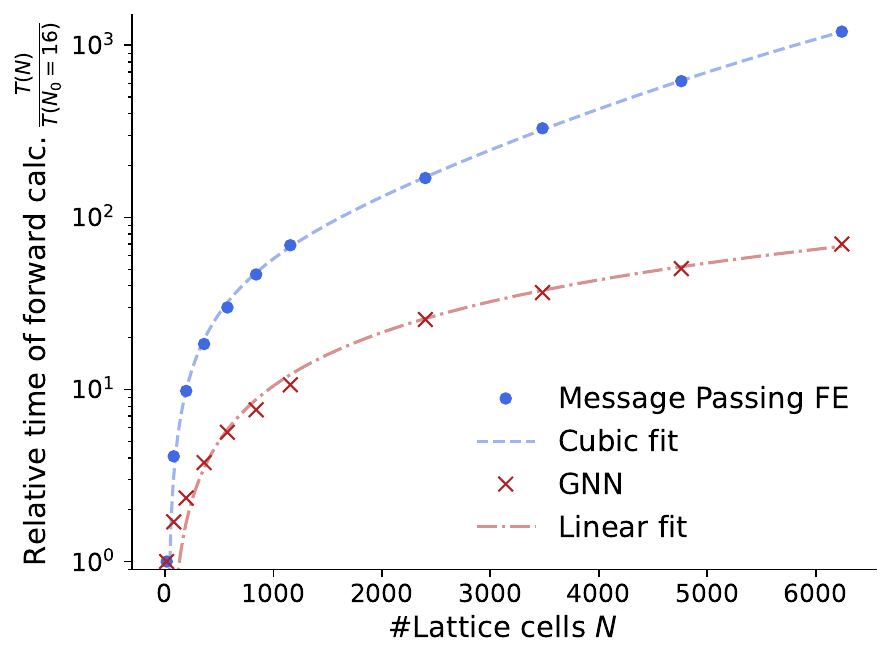}
	\caption{Relative time required for a single forward pass of the exact algorithm using message passing finite element (FE) and the approximate algorithm using GNNs. Experiments are done using square lattices. Time complexity is given in relation to the time required for a square lattice with $N_0 = 4 \times 4$ cells. 
	}
	\label{fig:Tcompl}
\end{figure}

The time complexity of the Message Passing Finite Element method can be estimated by looking at the three main steps of the algorithm: 
\begin{enumerate}
    \item Assembling the stiffness matrix, which scales with the number of edges $\mathcal{O}(|\mathcal{E}|)$.
    \item Solving for displacements, which scales, e.g. for algorithms like LU decomposition, cubic with the number of nodes $\mathcal{O}(k \cdot |\mathcal{X}|^3)$, where $k$ is the number of load iterations.
    \item The backward path, i.e., automatic differentiation, which scales proportional to the forward calculation (steps 1 and 2), with a low constant pre-factor\cite{baur1983complexity,griewank1989automatic} on the order of $\mathcal{O}(1)$.
\end{enumerate}
All in all, the algorithm scales approximately $\mathcal{O}(|\mathcal{E}|+k \cdot |\mathcal{X}|^3)$, with $k$ usually being $\mathcal{O}(10)$.
In our case, the number of cells $N$ in the lattice scales linearly with the number of edges or number of nodes, and thus we get $\mathcal{O}(k \cdot N^3)$.
This is consistent with experimental measurements, as shown in \cref{fig:Tcompl}.

In contrast, the time complexity of the GNN-based approach only scales, regarding lattice size, with the number of edge operations that have to be performed, therefore yielding a linear dependence on the number of cells, $\mathcal{O}(N)$, also observed in \cref{fig:Tcompl}.

\subsection{Finding most similar lattice designs in the training data}\label{SI:GNNdesign}

\begin{figure}[htb!]
    \centering
    \includegraphics[width=0.45\columnwidth]{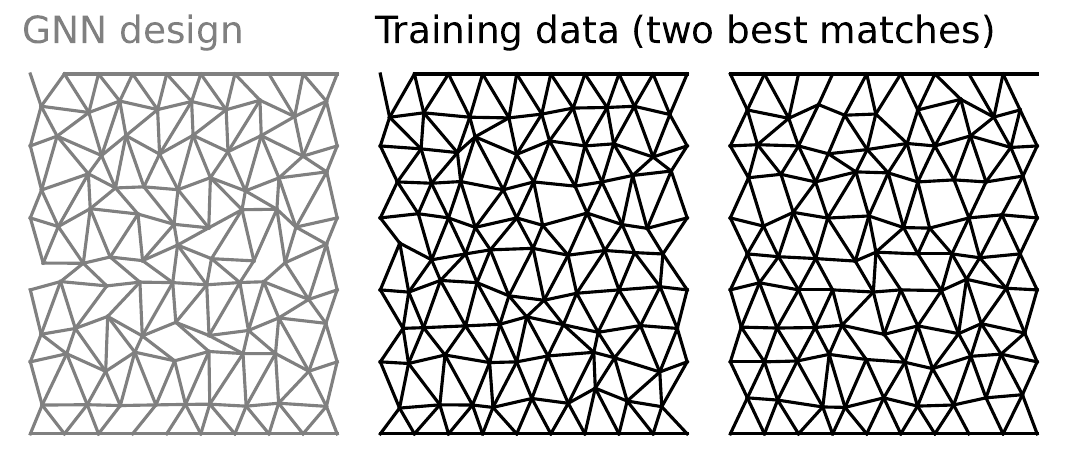}
	\caption{
	Images of the two lattices in the training set that resemble the found design most. Related to \cref{fig:invDesign_GNN}B.
	}
\label{fig:invDesignTrain}
\end{figure}

We evaluate the similarity between lattice designs using their image representation.
The 2D image of the generated lattice design is denoted by $\pmb{\mathcal{I}}_\text{GNN}$, while the images of the lattices in the training dataset are denoted by $\pmb{\mathcal{I}}_i$.
Dissimilarity $\Delta_\text{s}$ is calculated using
\begin{subequations}
\begin{align}
    \Delta_{\text{s}1}(\pmb{\mathcal{I}}_\text{GNN}, \pmb{\mathcal{I}}_i) &= \frac{\text{mean}\left(\left(\pmb{\mathcal{I}}_\text{GNN} \lor \pmb{\mathcal{I}}_i \right) \land \pmb{\mathcal{I}}_\text{GNN}\right)}{\text{mean}(\pmb{\mathcal{I}}_\text{GNN})} \,, \\
    \Delta_{\text{s}2}(\pmb{\mathcal{I}}_\text{GNN}, \pmb{\mathcal{I}}_i) &= \frac{\text{mean}\left(\left(\pmb{\mathcal{I}}_\text{GNN} \lor \pmb{\mathcal{I}}_i \right) \land \pmb{\mathcal{I}}_i \right)}{\text{mean}(\pmb{\mathcal{I}}_i)} \,, \\
    \Delta_\text{s}(\pmb{\mathcal{I}}_\text{GNN}, \pmb{\mathcal{I}}_i) &=\Delta_{\text{s}1}(\pmb{\mathcal{I}}_\text{GNN}, \pmb{\mathcal{I}}_i) + \Delta_{\text{s}2}(\pmb{\mathcal{I}}_\text{GNN}, \pmb{\mathcal{I}}_i)
\end{align}
\end{subequations}
where $\text{mean}$ calculates the mean over all pixels and we treat Boolean values as integers (False -- 0; True -- 1) when taking the mean.
For the experiment presented in \cref{fig:invDesign_GNN}B, the two lattices in the training dataset with the lowest dissimilarity to the found inverse design (as defined here) are shown in \cref{fig:invDesignTrain}.

\subsection{Open source package \em{pyLattice2D}}\label{SI:pyLattice2D}

pyLattice2D \cite{pylattice} is a Python package that implements an end-to-end differentiable framework for performing finite element analysis and inverse design of lattice materials in pyTorch.
It contains functions for generating a variety of lattices (Square, Equilateral Triangle, Honeycomb, Reentrant Honeycomb, Kagome and Voronoi) with (i) different number of cells and (ii) custom deformations such as node displacements and edge (or node) deletions.
In addition, code for training GNNs to predict lattice properties as well as various example notebooks for inverse design problems are included.

In the code, the direct stiffness matrix is constructed in \verb|fem_solver/direct_stiffness.py|, with convenience classes available for setting constraints in \verb|fem_solver/constraints_and_deformations.py|.
Code containing the logic for generating different base lattices is in \verb|lattices/|, while the pipeline for dataset generation is in \verb|data/create.py|.
The experimental protocol for obtaining the effective elastic modulus and Poisson's ratio of a lattice is implemented in \verb|methods/mechanical_properties.py|.
Wrapped models to be used are found in \verb|models/|, with \verb|models/Lattice.py| implementing the class describing lattices as graphs.
\verb|models/FEM.py| is the differentiable finite element solver, taking the aforementioned lattice objects as input and performing a single finite element step.
\verb|models/MPNN/| contains both GNN models as well as tabular and image-based machine learning models for property prediction.
Finally, \verb|Examples/| in the main folder features IPython notebooks with the experiments from this paper, and \verb|Data_Generation| features the Python and Bash scripts used to generate the dataset.\\

\noindent \textbf{\Large{Supplemental figures}}\\

\begin{figure}[htb!]
    \centering
    \includegraphics[width=0.5\textwidth]{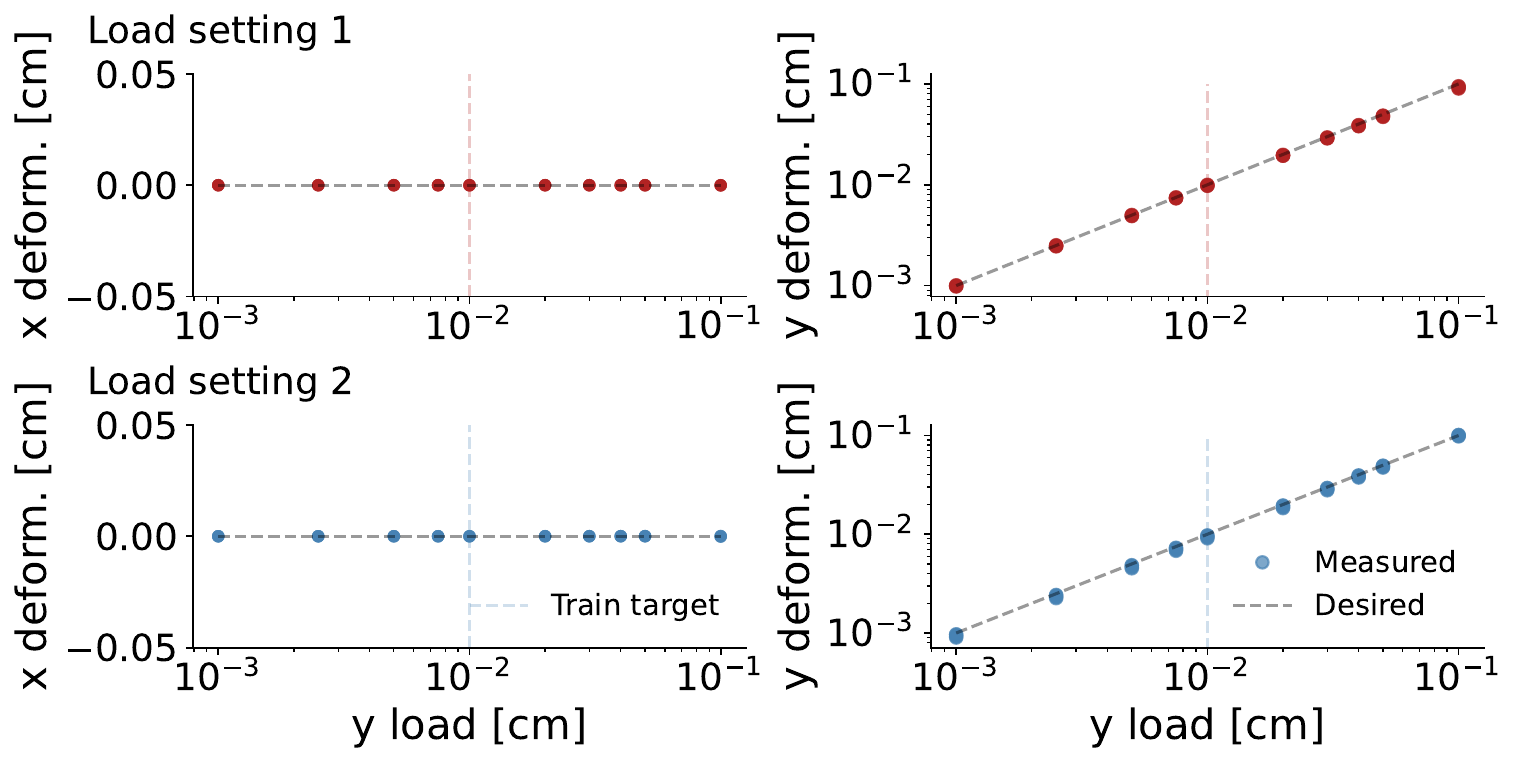}
	\caption{
	Target deformation (horizontal dashed line) and observed deformation (dots) for different absolute load strengths. Even for loads that are much lower or higher than the one used for training (vertical dashed line), the found lattice material obeys the desired target behaviour. Related to \cref{fig:invDesign_Flat}B.
	}	\label{fig:invDesign_FlatSI}
\end{figure}

\begin{figure}[ht!]
    \centering
    \includegraphics[width=0.55\textwidth]{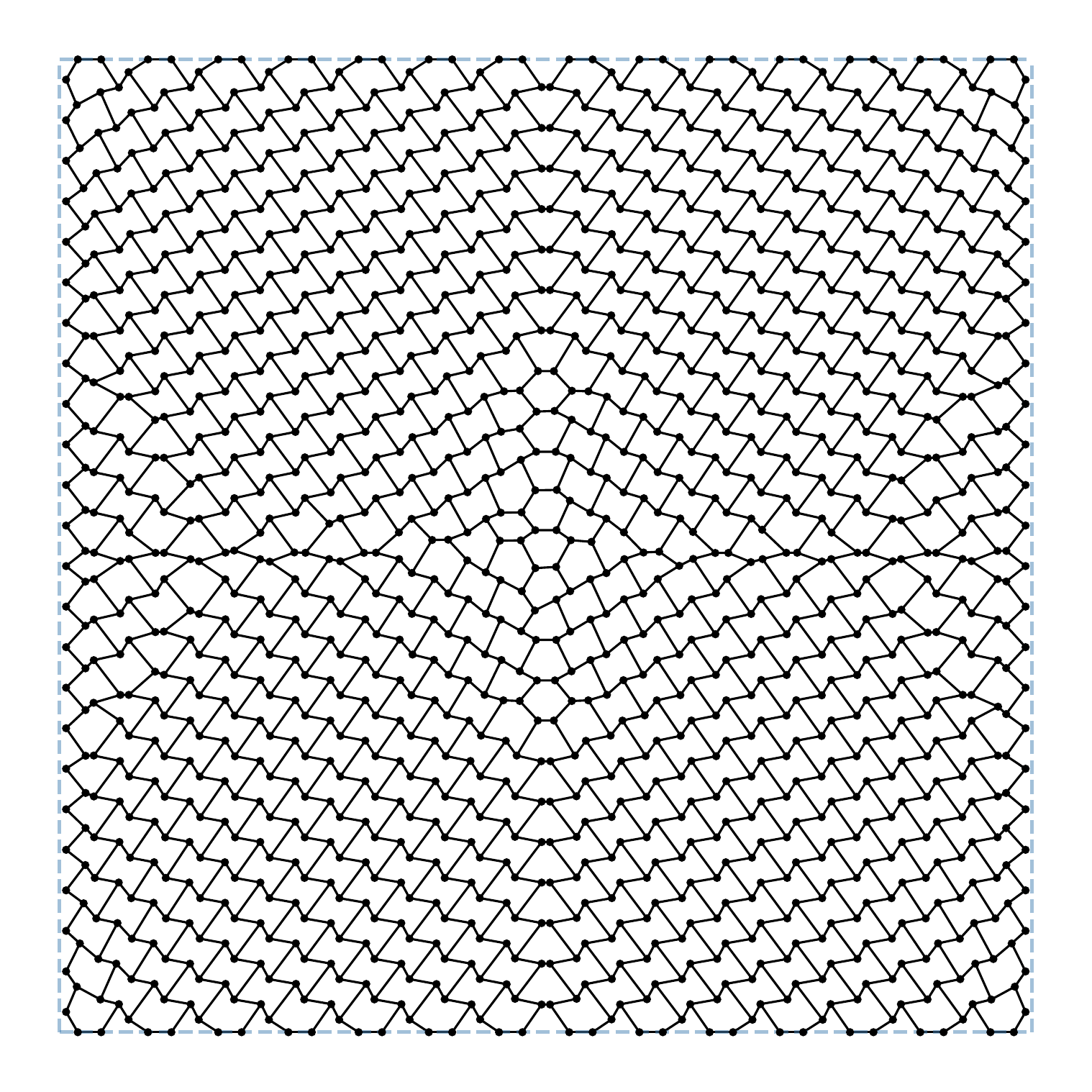}\vspace{-5mm}
	\caption{Magnified illustration of the inverse-designed lattice shown in \cref{fig:Large}C. Connection points between beams are shown as nodes. It should be noted that, although the design looks symmetric on first glance, it is not.
	}	
	\label{fig:LargeAuxMagnified}
\end{figure}

\begin{figure*}[htb!]
    \centering
    \includegraphics[width=0.95\textwidth]{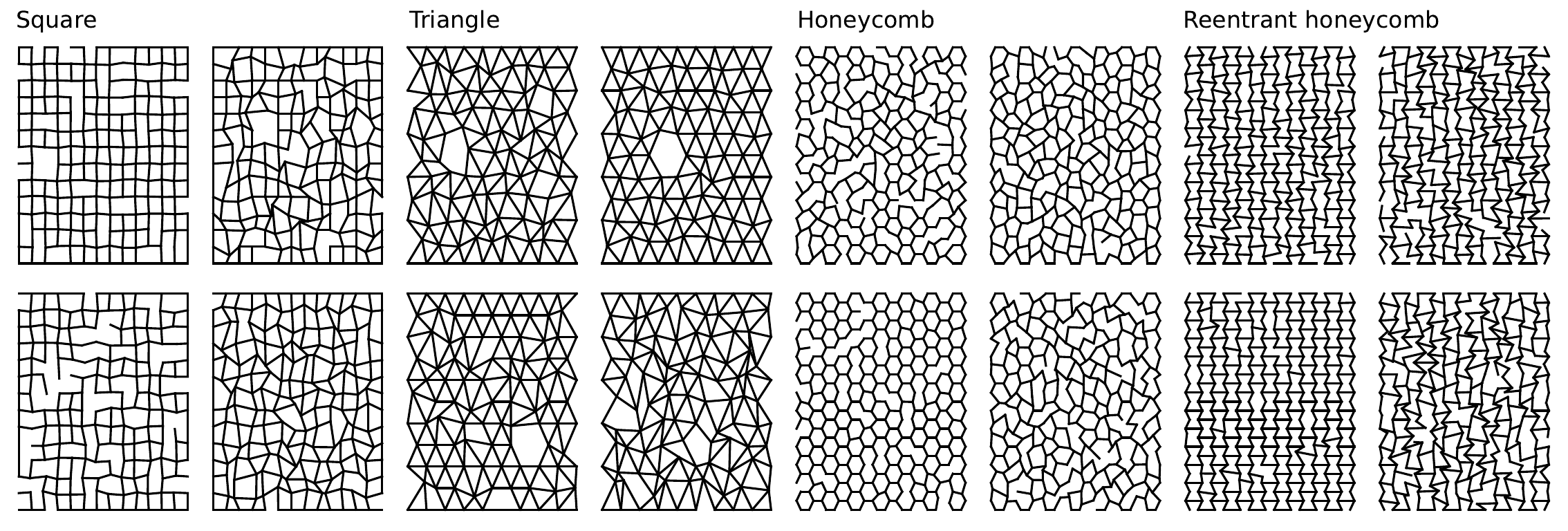}
	\caption{
	Example lattices from the generated training dataset used to train GNNs. Related to \cref{fig:prediction,fig:invDesign_GNN} and \cref{tab:results}.
	}
	\label{fig:dataset}
\end{figure*}

\begin{figure}[h!]
    \centering
    \includegraphics[width=0.55\textwidth]{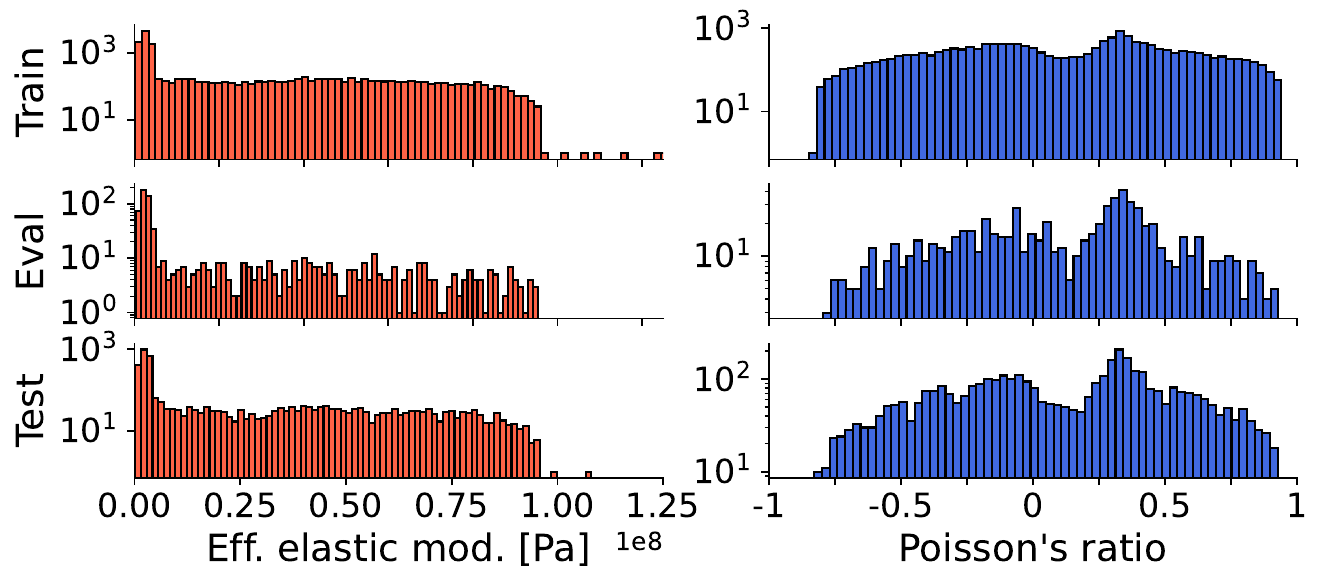}
	\caption{
	Distribution of effective elastic modulus and Poisson's ratio as obtained from our dataset, shown for the different data splits. The abscissas use log-scaling with different ranges. Related to \cref{fig:dataset}.
	}
	\label{fig:dstatistics}
\end{figure}

\FloatBarrier

\noindent \textbf{\Large{Supplemental tables}}\\

\begin{table}[htb!]\renewcommand{\arraystretch}{1.5}
\centering
\begin{tabular}{c||c|c|c}
\hline\hline
\begin{tabular}[c]{@{}c@{}}Lattice type\\ (2D)\end{tabular}       
& \begin{tabular}[c]{@{}c@{}}Relative density\\ $\left(\dens\right)$\end{tabular} &
\begin{tabular}[c]{@{}c@{}}Eff. elastic modulus\\ ($E^*$)\end{tabular} &
\begin{tabular}[c]{@{}c@{}}Poisson's ratio\\ ($\nu^*$)\end{tabular} \\ 
\hline\hline
Square    & $2\frac{t}{L}$ & $\frac{1}{2} \bar\rho \Es$  & 0                 \\ \hline
\begin{tabular}[c]{@{}c@{}}Triangular\\ (equilat.)\end{tabular}& $2\sqrt{3}\frac{t}{L}$ & $\frac{1}{3} \bar\rho \Es$  & $\frac{1}{3}$\\ \hline
Hexagonal & $\frac{2}{\sqrt{3}}\frac{t}{L}$ 
& $\frac{3}{2} \bar\rho^3 \Es$ & 1 \\ \hline
Reentrant & $\frac{8}{3 \sqrt{3}} \frac{t}{L}$ & $\frac{81}{128} \bar \rho^3 \Es$ & -1 \\[1mm]
\hline\hline
\end{tabular}\vspace{2mm}
\caption{Analytical relative density, effective elastic modulus and Poisson's ratio for various common 2D lattice grids.}\vspace{-4mm}\label{tab:rels}
\end{table}

\begin{table}[!htb]\renewcommand{\arraystretch}{1.2}
\begin{center}
\begin{tabular}{c | c c  c c c}
\hline\hline
Lattice type (2D) & $N_\text{l}$ & $N_\text{c}$ & $\Delta$ & $D_\text{n}$ & $D_\text{e}$ \\
\hline
Square & 11 & 169 & 0.15 & 1 & 0.2\\
Triangle & 14 & 170 & 0.15 & 1 & 0.2\\
Honeycomb & 25 & 150 & 0.05 & 0 & 0.2\\
Reentrant & 25 & 149 & 0.05 & 0 & 0.2\\
\hline\hline
\end{tabular}
\caption{Parameters used for dataset generation. Related to \cref{fig:invDesign_GNN} and \cref{tab:results}.}
\label{SI:dataTable}
\end{center}
\end{table}

\begin{table}[h!]\renewcommand{\arraystretch}{1.85}
\begin{tabular}{lc|lc}
\hline\hline
Feature & Formula & Feature & Formula \\ 
\hline
Image mean & $\mathrm{mean}(\pmb{\mathcal{I}})$ 
 & Beam length mean & $\mathrm{mean}(L)$ \\
Image standard deviation & $\mathrm{std}(\pmb{\mathcal{I}})$ & Beam length standard deviation & $\mathrm{std}(L)$ \\
Relative density & $\bar{\rho}$ & Beam length min/max & $\mathrm{min}(L)$, $\mathrm{max}(L)$ \\
Cell area mean & $\frac{1 - \mathrm{mean}(\pmb{\mathcal{I}})}{N_\text{c}}$ &
Cell area standard deviation & $\frac{\mathrm{std}\left(1 - \pmb{\mathcal{I}}\right)}{N_\text{c}}$ \\
\hline\hline
\end{tabular}
\centering
\caption{Tabular features for machine learning. 
$N_\text{c} = N_\text{b} - N_\text{n} + 1$ is the number of closed cells in the lattice, $N_\text{b}$ the number of beams and $N_\text{n}$ the number of nodes. Related to \cref{tab:results}.}\label{SI:dgeneration}
\label{tab:tabular}
\end{table}

\noindent \textbf{\Large{Supplemental videos}}\\

\noindent \textbf{Video S1}\\[-10pt]

Animation of the inverse design process for the grabber tool. The grabber is shown under load condition. During the course of inverse design iterations, the four nodes that are supposed to do the grabbing motion get closer to their target deformations, indicated by crosses. Deformations were magnified by a factor of $5$ here to ease visibility.\\

\noindent \textbf{Video S2}\\[-10pt]

Animation of the inverse design process for the flat surface, shown under both load conditions. Deformations were magnified by a factor of $8$ here to ease visibility.\\

\noindent \textbf{Video S3}\\[-10pt]

Visualisation of the hexagonal honeycomb lattice before (left) and after (right) inverse design, under a strong load.
We show all intermediate steps of deformation.
As expected, the hexagonal honeycomb lattice bulges outwards.
In contrast, the found design stays perfectly inside the bounding box.\\
\newrefcontext[labelprefix=S]
\renewcommand*{\bibfont}{\normalsize}
% \clearpage
\vspace{2mm}

\noindent \textbf{\Large{Supplemental references}}\\
\printbibliography[prefixnumbers=S, keyword=S,heading=none]%,heading=subbibintoc]
\end{refsection}
\end{document}